\documentclass[aps,prd,preprintnumbers,superscriptaddress,tighten,nofootinbib]{revtex4}
\usepackage{amssymb}
\usepackage{amsmath,color}
\usepackage{epsfig}
\usepackage{graphicx,epsf,epsfig}
\usepackage{bbm}
\def\slc#1{\setbox0=\hbox{$#1$}           
    \dimen0=\wd0                                 
    \setbox1=\hbox{/} \dimen1=\wd1               
    \ifdim\dimen0>\dimen1                        
       \rlap{\hbox to \dimen0{\hfil/\hfil}}      
       #1                                        
    \else                                        
       \rlap{\hbox to \dimen1{\hfil$#1$\hfil}}   
       /                                         
    \fi}

\DeclareMathOperator{\diag}{diag}

\begin{document}

\title{Signatures of Extra Dimensional Sterile Neutrinos}
\author{Werner Rodejohann}
\email{werner.rodejohann@mpi-hd.mpg.de}

\author{He Zhang}
\email{he.zhang@mpi-hd.mpg.de}

\affiliation{Max-Planck-Institut f\"ur Kernphysik, Saupfercheckweg 1, 69117 Heidelberg, Germany}

\begin{abstract}
\noindent
We study a large extra dimension model with active and sterile Dirac neutrinos. The sterile
neutrino masses stem from compactification of an extra dimension with radius $R$
and are chosen to have masses around eV or keV, in order to explain short-baseline 
anomalies or act as
warm dark matter candidates.
We study the effect of the sterile neutrino Kaluza-Klein tower in short-baseline 
oscillation experiments and in the beta spectrum as measurable by KATRIN-like experiments.
\end{abstract}

\maketitle
\section{Introduction}
\label{sec:intro}

Remarkable experimental activity in the past decades has established that the phenomenon
of neutrino flavor transition is described by standard three neutrino oscillations.
There are however experimental hints \cite{Mention:2011rk,Huber:2011wv}, mainly by short-baseline
experiments, towards the existence of sterile neutrinos, which do not participate in ordinary weak interactions (see Ref.~\cite{Kopp:2013vaa} for a recent global fit).
In addition, several hints mildly favoring extra radiation in the Universe have emerged from precision cosmology and Big Bang Nucleosynthesis~\cite{Cyburt:2004yc,Izotov:2010ca,Hamann:2010bk,Giusarma:2011ex}.
Taking the recently claimed detection of $B$-mode polarization from the BICEP2 experiment~\cite{Ade:2014xna} into account increases the significance~\cite{Giusarma:2014zza}.

Apart from the hypothesis of such light sterile neutrinos, a keV scale sterile neutrino, which is the front running candidate for warm dark matter (WDM), draws a lot of attention. Compared to cold dark matter, a WDM candidate provides a smoother inner mass density profile of dark matter halos, which is generally more cuspy in cold dark matter simulations. Another advantage of WDM is that the observed small scale structure can be reproduced, whereas the predicted number of satellite galaxies within cold dark matter scenarios typically exceeds observation. There was also an interesting indication of keV WDM from the stacked X-ray spectrum of galaxies and clusters~\cite{Bulbul:2014sua,Boyarsky:2014jta}, which points to a sterile neutrino with mass $m_s \simeq 7.1~{\rm keV}$ and mixing angle $\sin^2\theta_s \sim 10^{-11}$.

While none of the hints is fully convincing, it is worthwhile to speculate on the possible origin of  sterile neutrinos and their mass scale. Indeed, a lot of models have been built with the attempt of embedding sterile neutrinos in more fundamental frameworks. Possibilities include theories with extra dimensions,  exponentially suppressing fermion masses e.g.\ by localizing them on a distant brane. This has been proposed to generate seesaw neutrinos of keV scale in~\cite{Kusenko:2010ik}, see also~\cite{Adulpravitchai:2011rq}. Flavor symmetries~\cite{Altarelli:2010gt,Ishimori:2010au} can predict that one of the heavy neutrino masses is zero. Slightly breaking this symmetry generates a neutrino with much smaller mass than the other two, whose masses are allowed by the symmetry. This has been proposed to generate seesaw neutrinos of keV scale in~\cite{Mohapatra:2005wk,Shaposhnikov:2006nn,Lindner:2010wr}. While the commonly studied flavor models with non-abelian discrete symmetries cannot produce a non-trivial hierarchy between fermion masses, the Froggatt-Nielsen mechanism is capable of this~\cite{Froggatt:1978nt} and has been proposed to generate seesaw neutrinos of eV or keV scales in~\cite{Barry:2011fp,Merle:2011yv,Barry:2011wb}. Extensions or variants of the canonical type I seesaw often contain additional mass scales, which can be arranged to generate keV-scale particles~\cite{Barry:2011wb,Zhang:2011vh,Heeck:2012bz,Dev:2012bd}.

In this paper we study a large extra dimension model~\cite{ArkaniHamed:1998rs,Antoniadis:1998ig,Davoudiasl:2002fq}, in which the right-handed neutrinos are located on the 5D bulk, while all the other Standard Model particles are confined to the 4D brane. The compactification radius of our framework is chosen to be around $R^{-1} \simeq {\rm eV}$ or ${\rm keV}$. After expanding the five-dimensional bulk neutrinos in Kaluza-Klein (KK) modes, the couplings between bulk and brane neutrinos lead to a Dirac mass term for active neutrinos as well as an admixture between active and heavier KK modes, which are the sterile neutrino states in our model\footnote{See \cite{Gingrich:2009az} for general constraints on such scenarios and \cite{Eingorn:2010wi} for a critical point of view.}. By imposing the condition of lepton number conservation, we find strong correlations between the
active neutrino parameters and the active-sterile neutrino mixing. Phenomenological consequences are also expected at near future neutrino facilities. We study effects of the setup in short-baseline neutrino oscillation experiments and in beta decays, pointing out the effect of the KK tower in particular. 

The outline of the paper is as follows: In Sec.~\ref{sec:model} we present the extra dimension model and its characteristic features. In Sec.~\ref{sec:parameters} we discuss the model parameters and the predictions on the active-sterile mixing. In Sec.~\ref{sec:pheno} we focus on the signatures of the model at current and future neutrino facilities, in particular short-baseline neutrino oscillation experiments and KATRIN.

\section{The model} \label{sec:model}

We consider a brane world theory with a five-dimensional bulk, where
the SM particles are confined to the brane. We also introduce three
SM singlet fermions $\Psi_i$ ($i=1,2,3$)
\cite{Dienes:1998sb,ArkaniHamed:1998vp,Dvali:1999cn,Barbieri:2000mg,Lukas:2000rg,BastoGonzalez:2012me}.
Being singlets, they are not restricted to the brane and can
propagate in the extra spacetime dimensions. The action responsible
for the neutrino masses is given by
\begin{eqnarray}\label{eq:S5}
S  =  \int {\rm d}^4 x {\rm d}y \left[ {\rm i}\overline{\Psi}
\slc{D} \Psi - \frac{1}{2} \left(\overline{\Psi^c} M_{\rm R} \Psi +
{\rm h.c.} \right) \right] 
   +\int_{y=0} {\rm d}^4 x \left( - \frac{1}{\sqrt{M_S}}
\overline{\nu_{\rm L}} \hat m^c \Psi - \frac{1}{\sqrt{M_S}}
\overline{\nu^c_{\rm L}} \hat m \Psi + {\rm h.c.}\right),
\end{eqnarray}
where $y$ is the coordinate along the extra compactified dimension
and $M_S$ denotes the mass scale of the higher-dimensional theory. 
The Dirac masses $\hat m$ and $\hat m^c$ could be generated by couplings
of the bulk neutrinos to a brane-localized Higgs boson 
after electroweak symmetry breaking. 
Note that, although $\Psi^c$ is defined in the same way as in four
dimensions, it does not represent the charge conjugate of $\Psi$ in
five dimensions \cite{Pilaftsis:1999jk}, and hence, the term
$\overline{\Psi^c} M_{\rm R} \Psi$ is not a Majorana mass
term\footnote{Majorana mass terms are not allowed in
five-dimensional spacetime \cite{Weinberg:1984vb}.}. However, in the
four-dimensional theory, it leads to effective Majorana mass terms
for the KK modes of $\Psi$. In this work we will for simplicity and definiteness
assume the conservation of lepton number and take $M_{\rm R}=0$. 

We decompose the spinors of the bulk singlet fermions into two
two-component objects: $\Psi=(\xi \,,\, \eta^c)^T$, where $\eta^c =
{\rm i} \sigma^2 \eta^*$. Since the extra dimension is compactified
on a $S^1/\mathbb{Z}_2$ orbifold, the KK modes of $\xi$ and
$\eta^c$ are four-dimensional Weyl spinors. We take $\xi$ to be even
under the ${\mathbb Z}_2$ transformation $y\to-y$, while $\eta$ is
taken to be odd. Thus, in Eq.~\eqref{eq:S5}, the $\hat m^c$ term
corresponding to the coupling between $\nu_L$ and $\eta$ is not
allowed. The KK expansions of $\xi$ and $\eta$ are given by
\begin{eqnarray}\label{eq:expand}
\xi(x,y) & = & \frac{1}{\sqrt{\pi R}}\xi_{(0)}(x) +
\sqrt{\frac{2}{\pi R}} \sum^N_{n=1} \xi_{(n)}(x)
\cos\left(\frac{ny}{R}\right), \nonumber \\
\eta(x,y) & = & \sqrt{\frac{2}{\pi R}} \sum^N_{n=1} \eta_{(n)}(x)
\sin\left(\frac{ny}{R}\right).
\end{eqnarray}
Inserting the above expansion into Eq.~\eqref{eq:S5} and integrating
over the compactified extra dimension, we arrive at the following
form of the four-dimensional action
\begin{eqnarray}\label{eq:S4}
S & = & \int {\rm d}^4 x \left\{{\xi}^{\dagger}_{(0)} {\rm i}
\bar{\sigma}^\mu
\partial_\mu \xi_{(0)} +  \sum^N_{n=1} \left(
{\xi}_{(n)}^{\dagger} {\rm i} \bar{\sigma}^\mu
\partial_\mu \xi_{(n)} +
{\eta}_{(n)}^{\dagger} {\rm i} \bar{\sigma}^\mu
\partial_\mu \eta_{(n)} \right) \right. \nonumber \\
&& \phantom{\int d^4 x}-\frac{\rm i}{2}\left[ \left.   \sum^N_{n=1} \left(
\begin{matrix} {\xi_{(n)}}^T & {\eta_{(n)}}^T
\end{matrix} \right) \sigma^2 {\cal M}_n \left( \begin{matrix} {\xi_{(n)}}
\cr {\eta_{(n)}} \end{matrix} \right) + {\rm h.c.} \right] \right.
\nonumber
\\
&& \phantom{\int d^4 x}-\left. {\rm i}  \left(\nu_{\rm L}^T \sigma^2
m_{\rm D} \xi_{(0)} + \sqrt{2} \sum^N_{n=1} \nu_{\rm L}^T \sigma^2
m_{\rm D} \xi_{(n)} + {\rm h.c.} \right) \right\},
\end{eqnarray}
where $\bar \sigma_\mu = ({\bf 1},\sigma_i)$ with $\sigma_i$ being the Pauli matrices. Written in block-form, the mass matrix ${\cal M}_n$ for the
KK modes at the $n$th level takes the form
\begin{eqnarray}\label{eq:MN}
{\cal M}_n = \left(\begin{matrix} 0 & n/R \cr  n/R & 0
\end{matrix}\right).
\end{eqnarray}
The Dirac mass term is then given by $m_{\rm D} = \hat m /\sqrt{2\pi
M_S R}$.
Because of the freedom in the choice of basis for the bulk
fermions, one can always apply a unitary transformation in flavor
space on $\xi_{(0)}$ and $\xi_{(n)}$ in order to make $m_{\rm D}$ hermitian. We therefore take
$$m_{\rm D}=U {\rm diag}(m_1,m_2,m_3) U^\dagger = U D U^\dagger\,,$$
with $U$ being a unitary matrix.

The full mass matrix in the basis $\left\{\nu_{\rm L},\xi_{(0)},
\xi_{(1)},\eta_{(1)}, \xi_{(2)},\eta_{(2)},\ldots\right\}$ then
reads
\begin{equation}\label{eq:M}
{\cal M} = \left(\begin{array}{cc|cc|ccc}
0 & m_{\rm D} & \sqrt{2} m_{\rm D} & 0 &  \sqrt{2} m_{\rm D}  & 0 & \cdots
\cr  m^T_{\rm D}  & 0  & 0 & 0  & 0 & 0 & \cdots
\cr  \hline \sqrt{2}  m^T_{\rm D} & 0 & 0 & \displaystyle \frac{1}{R} &   0 & 0 &  \cdots
\cr 0 & 0 & \displaystyle  \frac{1}{R}  & 0 & 0  & 0 &   \cdots
\cr  \hline \sqrt{2}  m^T_{\rm D}  &  0 & 0 & 0 & 0 & \displaystyle  \frac{2}{R}  &  \cdots
\cr 0 & 0 & 0 & 0 & \displaystyle  \frac{2}{R}  & 0 &  \cdots
\cr  \vdots  &  \vdots   &  \vdots   &  \vdots  &  \vdots   &  \vdots   &  \ddots
\end{array}\right).
\end{equation}
The zero mode $\xi_{(0)}^{c}$ can now be viewed as the right-handed
component of active neutrinos, and the Dirac mass of active
neutrinos is simply given by $m_{\rm D}$. One can check that the
higher KK modes do not contribute to active neutrino masses, and there
is no seesaw mechanism. Actually, the heavy KK modes $\xi_{(n)}$ and
$\eta_{(n)}$ form Dirac pairs, i.e.\
${\nu_{s,n}}=\left(\eta_{(n)},\xi^c_{(n)}\right)^T$ (for $n\geq 1$)
with masses being $n/R$. In this notation, the neutrino mass term
can be rewritten as
\begin{eqnarray}\label{eq:Mn}
\overline{\left(\nu,\nu_{s,1},\nu_{s,2},\cdots\right)_{\rm L}}
\left(\begin{matrix} m_{\rm D} & \sqrt{2}m_{\rm D} &
 \sqrt{2} m_{\rm D}  & \cdots  \cr 0 & 1/R
 & 0 & \cdots\cr 0 & 0 & 2/R & \cdots \cr \vdots & \vdots & \vdots & \ddots
\end{matrix}\right) \begin{pmatrix} \xi^c_{(0)} \cr \nu_{s,1} \cr \nu_{s,2} \cr \vdots \end{pmatrix}_{\rm
R} \; .
\end{eqnarray}
The active neutrinos are therefore Dirac particles (hence there will be no neutrinoless double
beta decay)
and their mass matrix is $m_{\rm D}$. As a leading order approximation, the sterile neutrino
states obtain masses as $n/R$ (for $n=1,2,3,\ldots$).

Since we are interested in the low-scale phenomena induced by light sterile neutrinos much smaller than the electroweak breaking scale, it is
convenient to define the Hermitian form of the neutrino mass matrix as ${\cal H}={\cal M}{\cal M}^\dagger$, i.e.
 \begin{eqnarray}\label{eq:H}
{\cal H}=
\left(\begin{matrix} (1+\sqrt{2}n)U D^2  U^\dagger & N_{(1)} M^2_{s,1} &
N_{(2)} M^2_{s,2} & \cdots  \cr N^\dagger_{(1)} M^2_{s,1} & M^2_{s,1}
 & 0 & \cdots\cr N^\dagger_{(2)} M^2_{s,2} & 0 & M^2_{s,2} & \cdots \cr \vdots & \vdots & \vdots & \ddots
\end{matrix}\right) ,
\end{eqnarray}
where $M_{s,n}=\diag(nR^{-1},nR^{-1},nR^{-1})$, and $N_{(n)} = \sqrt{2}n^{-1} U D U^\dagger R$. Up to order ${\cal O}(M^2_{\rm D}R^{-2})$, the mixing
matrix between active and $n$-th KK sterile neutrinos is approximately given by $K_{(n)} = \sqrt{2}n^{-1} U D R$. In the conventional notation, one can
write $U_{e4}=K_{(1)1,1}$, $U_{e5}=K_{(1)1,2}$, $U_{e6}=K_{(1)1,3}$ and so
on. Note that, since $K_{(n)}$ is inversely proportional to $n$, for heavier sterile neutrinos
the mixing is suppressed by $1/n$. We stress here the attractive feature of this 
scenario, namely that the mixing between active and sterile neutrinos is determined by 
measurable active neutrino parameters and the size of the extra dimension. 

The above mass matrix is given in the flavor basis, while the weak interaction Lagrangian for the leptons can now be written in the mass basis as
\begin{eqnarray}\label{eq:L}
{\cal L}_{\rm CC} & = & -\frac{g}{\sqrt{2}}   \ell^\dagger
\bar{\sigma}^\mu P_{\rm L}\left[ U  \nu_{m} + \sum^{\infty}_{n=1}
K_{(n)} \nu_{s,n}  \right]W^-_\mu + {\rm h.c.}, \\
{\cal L}_{\rm NC} & = & \frac{g}{2 \cos \theta_{\rm W}} {\nu_{m}^\dagger} \bar{\sigma}^\mu U^\dagger P_{\rm L}\left[ \sum^{\infty}_{n=1}
K_{(n)} \nu_{s,n} \right] Z_\mu+ {\rm h.c.},
\end{eqnarray}
where $\theta_W$ denotes the weak mixing angle. Due to the existence of sterile neutrinos, the $3\times 3$ active neutrino mixing
matrix is not unitary any more, and to a good approximation we have
\begin{eqnarray}\label{eq:MNS}
V & \simeq &    \left(1-\frac{1}{2}\sum^\infty_{n=1}
K_{(n)}{K_{(n)}^\dagger} \right)U
\end{eqnarray}
for active neutrino flavor mixing.

The active-sterile mixing depends strongly on the size of the extra dimension or the sterile neutrino mass. For example, if we take $R^{-1} \sim
1~{\rm eV}$ and $m_{\rm D} \sim 0.1 ~{\rm eV}$, the mixing between
active and the first KK sterile neutrinos would be around $0.1$,
which is appropriate to explain short-baseline anomalies. If we take
$R^{-1}$ to be $1~{\rm keV}$, the largest active-sterile mixing is
given by $m_{\rm D}R \sim 10^{-4}$, in the right ballpark for the
sterile neutrino warm dark matter hypothesis. For example, in the
Dodelson-Widrow scenario \cite{Dodelson:1993je},
i.e.\ production of WDM by neutrino
oscillations, if one assumes that sterile neutrino WDM with mass $M_s$ and mixing $\theta_s$ makes
up all the DM in the Universe, its abundance is given by
\begin{eqnarray}\label{eq:WDM}
\Omega \simeq 0.2 \left(\frac{\theta^2_s}{3\times 10^{-9}}\right) \left(\frac{M_s}{3~{\rm keV}}\right)^{1.8} \; .
\end{eqnarray}
This mechanism is gradually building up the WDM density when the
plasma in the early Universe produces active neutrinos from which
the sterile ones originate via their small mixing with the active
ones. Another popular mechanism (Shi-Fuller \cite{Shi:1998km})
requires a lepton asymmetry which can generate sterile neutrinos via
resonant oscillations. 
As the generation of WDM, or the nature of dark matter in general, is not known,
most of the following analysis of keV-scale neutrinos is independent of
warm dark matter generation. To explicitely check if our framework can generate the
necessary WDM density requires a complex analysis (e.g.\ the presence
of the KK tower needs to be taken into account) beyond the scope of
this analysis based on phenomenology in terrestrial experiments. 

Let us sum up the main features of our framework: for each active neutrino there
is a sterile neutrino, which in turn has a Kaluza-Klein tower of heavier copies with smaller mixing.
With $R^{-1}$ larger than active neutrino masses, each set of KK states is to good
precision degenerate in mass. This mass is given by $n/R$, and the mixing with active
neutrinos is given by $\sqrt{2}n^{-1} U D R$, where $D$ is a diagonal matrix containing the
active neutrino masses and $U$ the PMNS matrix.

\section{Model parameters}
\label{sec:parameters}

We continue to discuss the free parameters in the model. In general, the unitary matrix $U$ can be parametrized by three mixing angles ($\theta_{12}$,
$\theta_{13}$, $\theta_{23}$) and one Dirac type CP-violating phase $\delta$, i.e.
\begin{eqnarray}\label{eq:SP}
U = \left(\begin{matrix}c_{12} c_{13} & s_{12} c_{13} & s_{13}
e^{-{\rm i}\delta} \cr -s_{12} c_{23} - c_{12} s_{23} s_{13}e^{{\rm
i}\delta} & c_{12} c_{23} - s_{12} s_{23} s_{13}e^{{\rm i}\delta} &
s_{23} c_{13} \cr s_{12} s_{23} - c_{12} c_{23} s_{13}e^{{\rm
i}\delta} & -c_{12} s_{23} - s_{12} c_{23} s_{13}e^{{\rm i}\delta} &
c_{23} c_{13}
\end{matrix}\right)   ,
\end{eqnarray}
where $s_{ij} \equiv \sin \theta_{ij}$ and $c_{ij} \equiv \cos \theta_{ij}$ (for $ij = 12, 23, 13$). Apart from the CP phase, all the three mixing
angles have already been measured experimentally, and the latest global fit data for the mixing angles can be found in
Ref.~\cite{GonzalezGarcia:2012sz}
\begin{eqnarray}\label{eq:bound}
\sin^2\theta_{12} &=& 0.313^{+0.013}_{-0.012} \; , \nonumber \\
\sin^2\theta_{23} &=& 0.444^{+0.036}_{-0.031} \; , \\
\sin^2\theta_{13} &=& 0.0244^{+0.0020}_{-0.0019} \; . \nonumber
\end{eqnarray}
Strictly speaking, in the presence of sterile neutrinos, the above fitted mixing angles are polluted by the mixing between active and sterile
neutrinos.
We will ignore small corrections of sterile neutrinos to the above numbers, and make use of the data
in Eq.~\eqref{eq:bound} for our numerical analysis. The neutrino mass-squared differences
are also given in Ref.~\cite{GonzalezGarcia:2012sz} as $\Delta m^2_{21}= 7.45 \times 10^{-5} ~{\rm eV}^2$ and $|\Delta m^2_{31}|= 2.417 \times
10^{-3} ~{\rm eV}^2$.

\begin{figure}[!xt]
\begin{center}\vspace{-0.0cm}
\includegraphics[width=.4\textwidth]{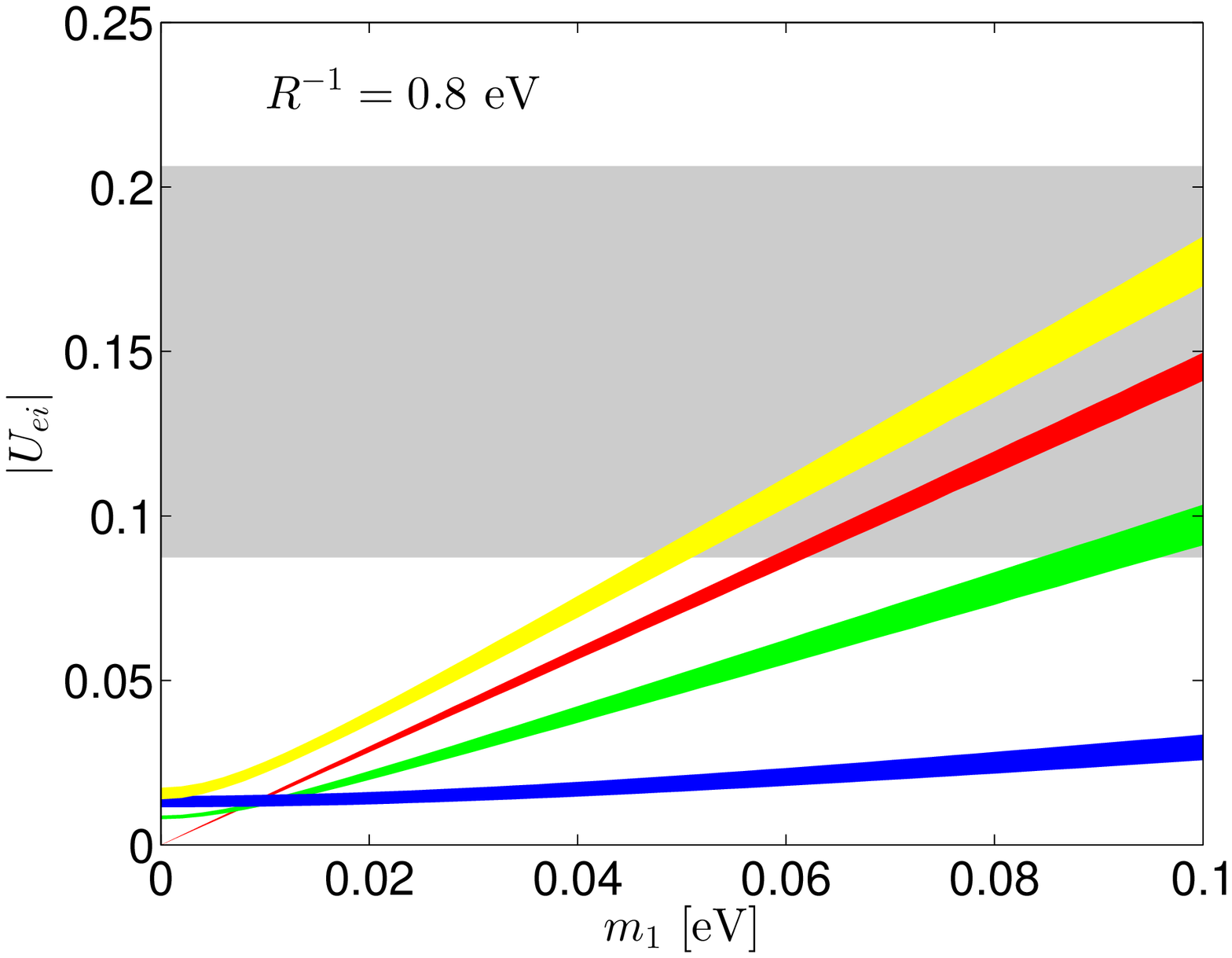}\hspace{5mm}
\includegraphics[width=.4\textwidth]{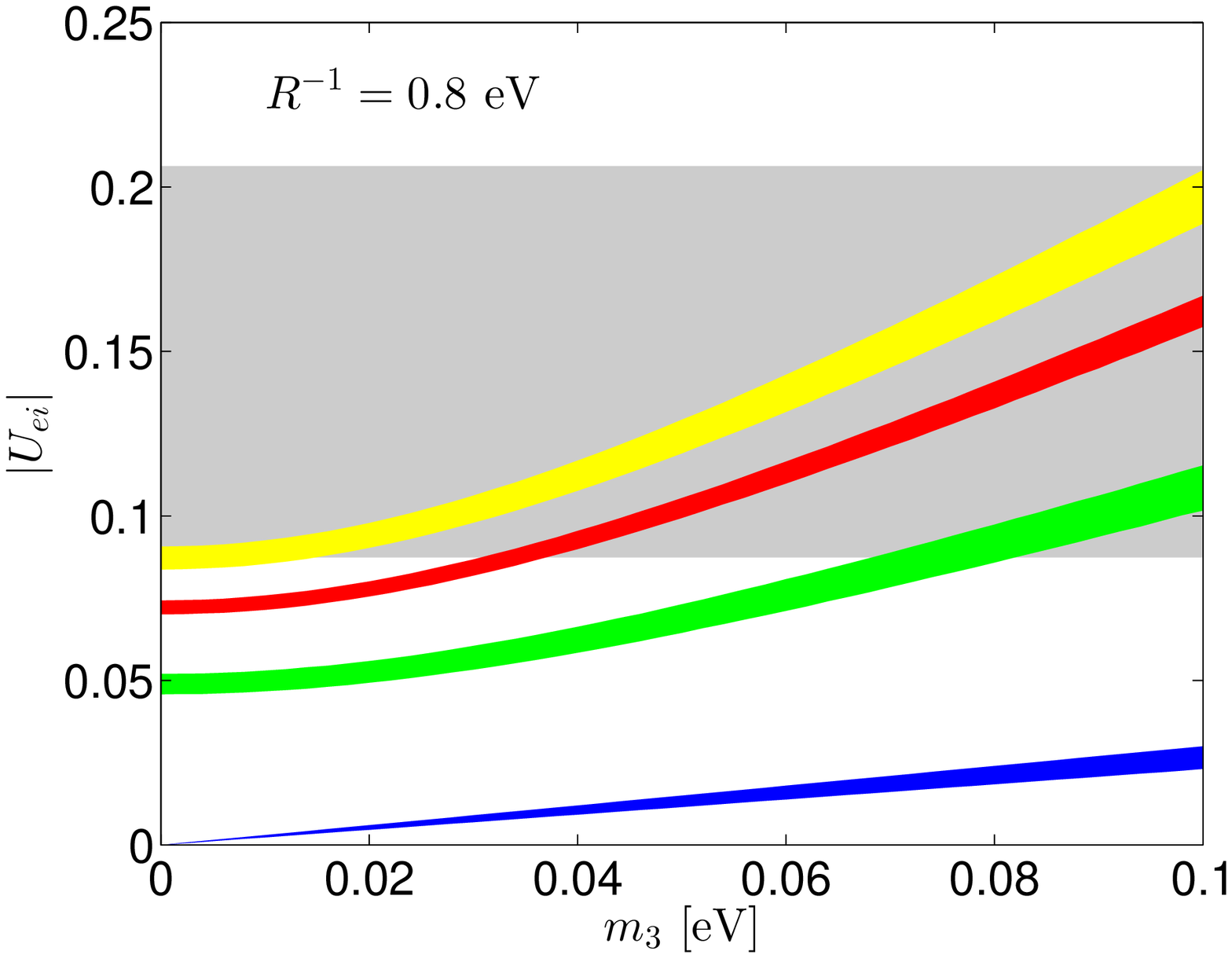}
\includegraphics[width=.4\textwidth]{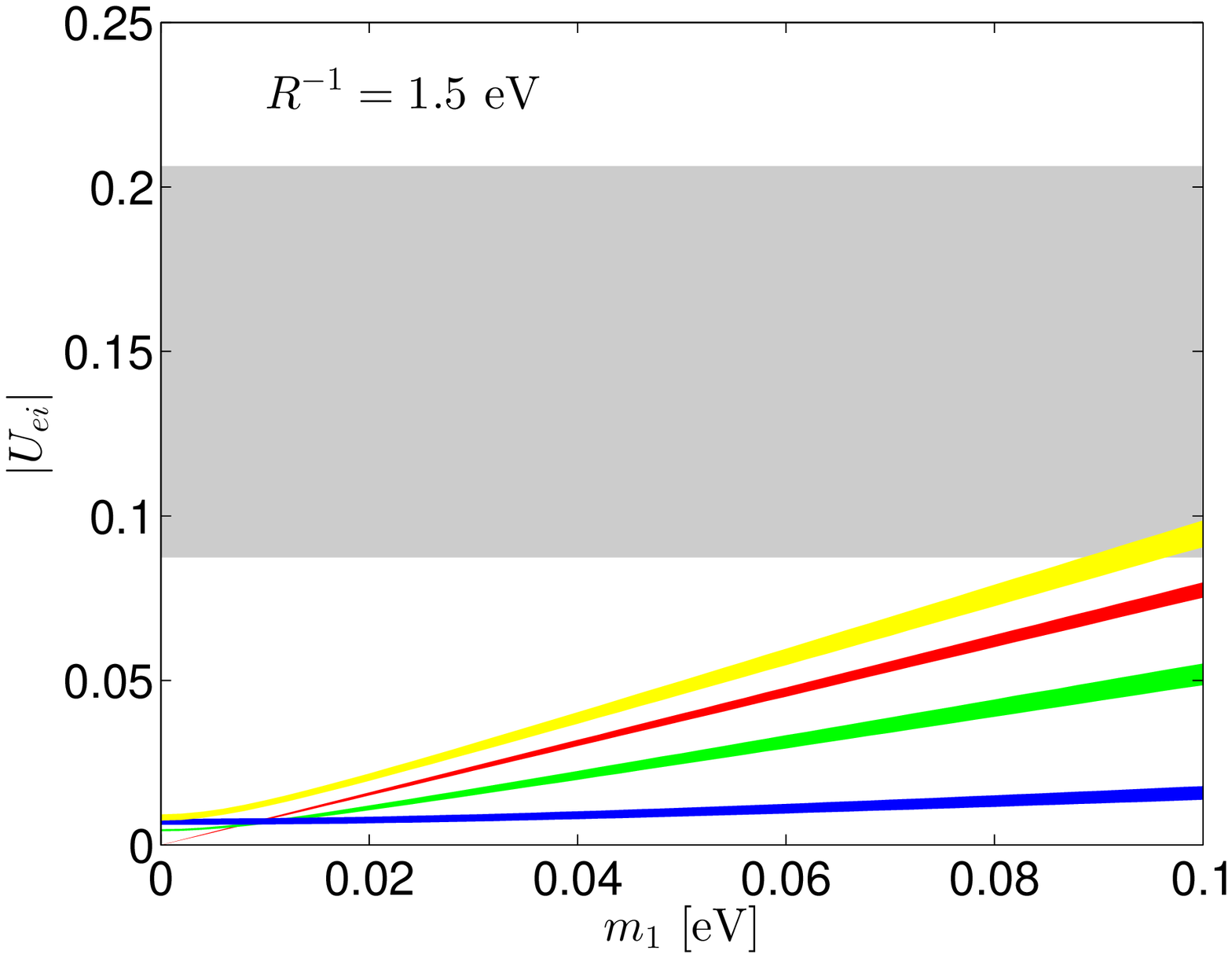}\hspace{5mm}
\includegraphics[width=.4\textwidth]{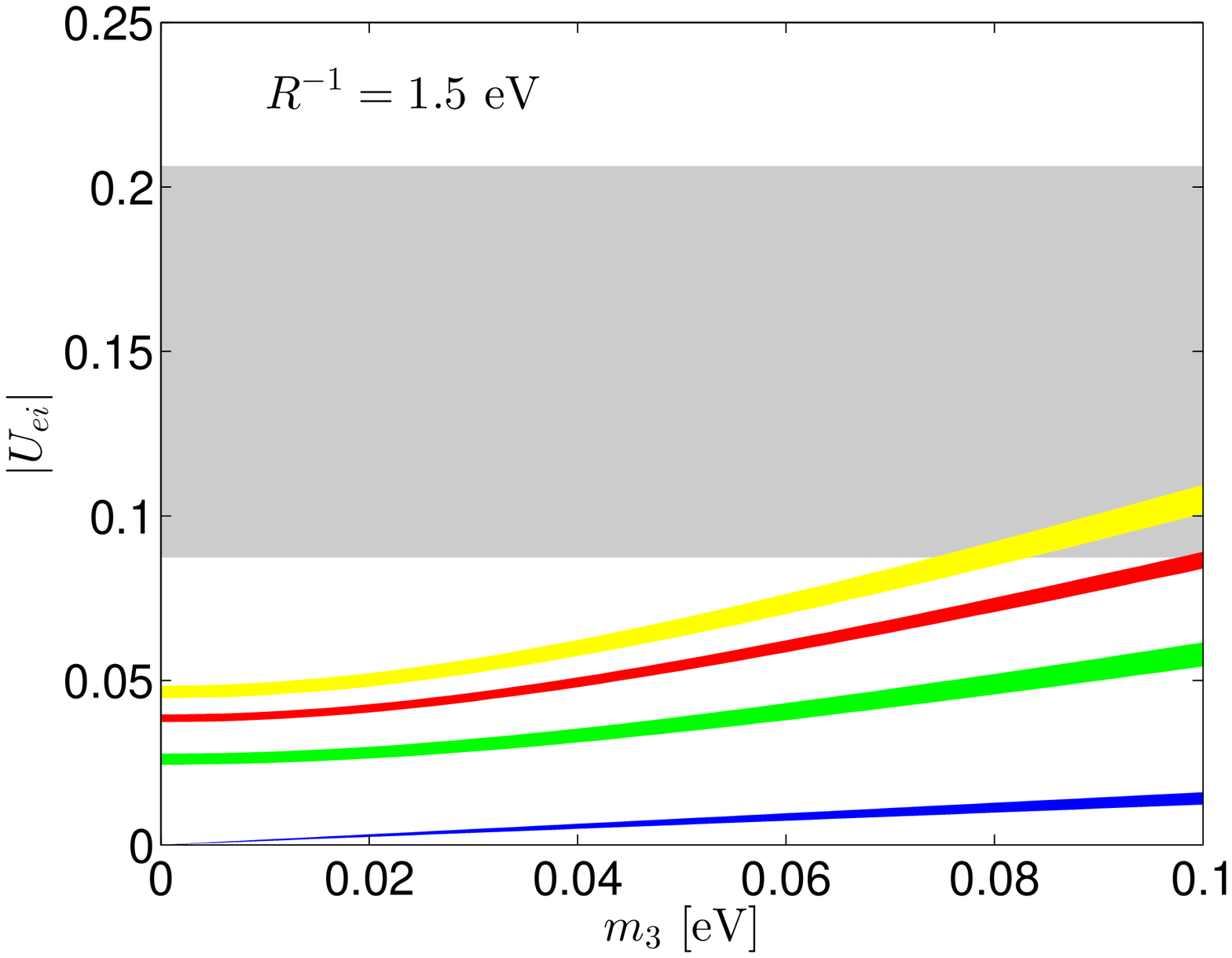}
\includegraphics[width=.4\textwidth]{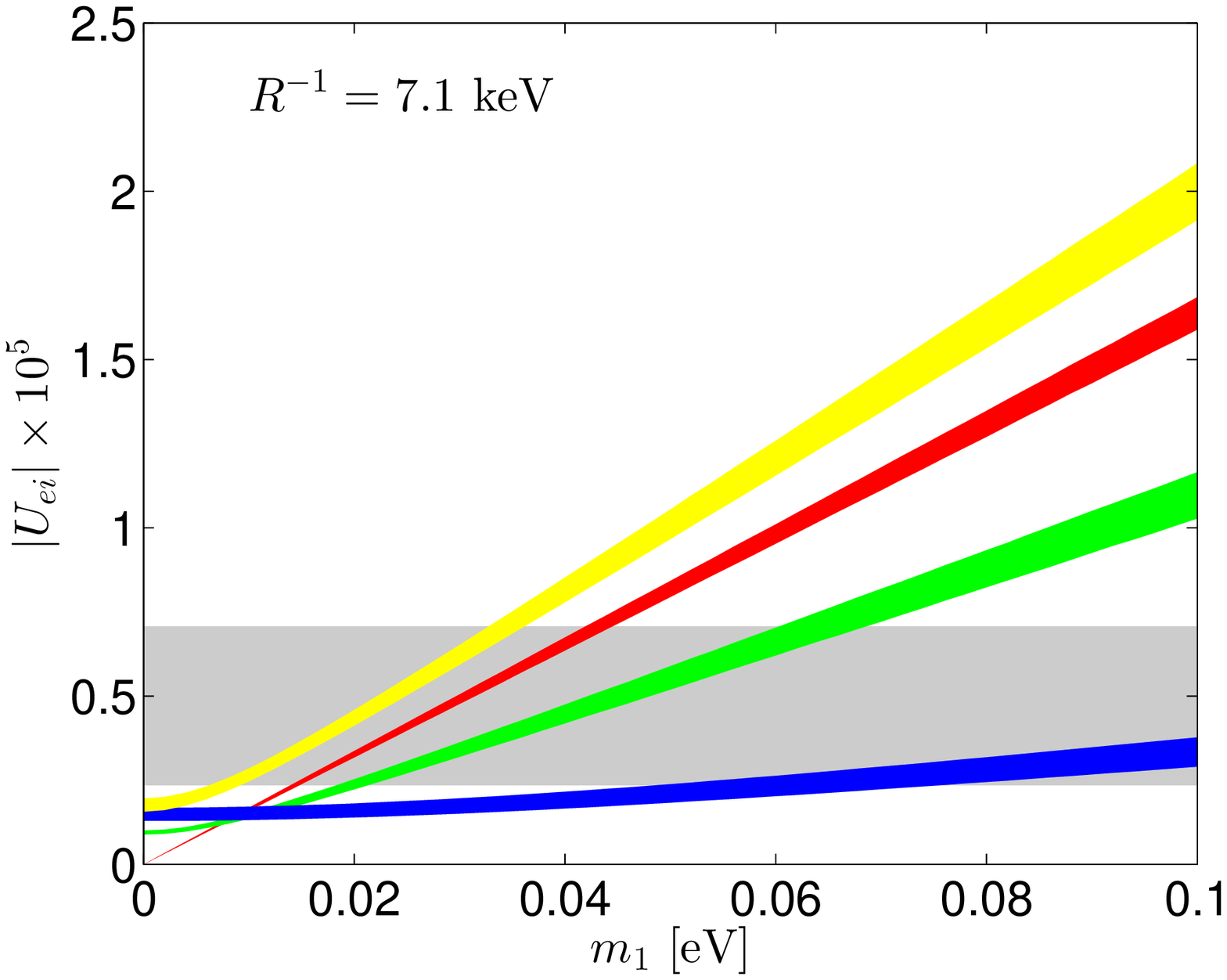}\hspace{5mm}
\includegraphics[width=.4\textwidth]{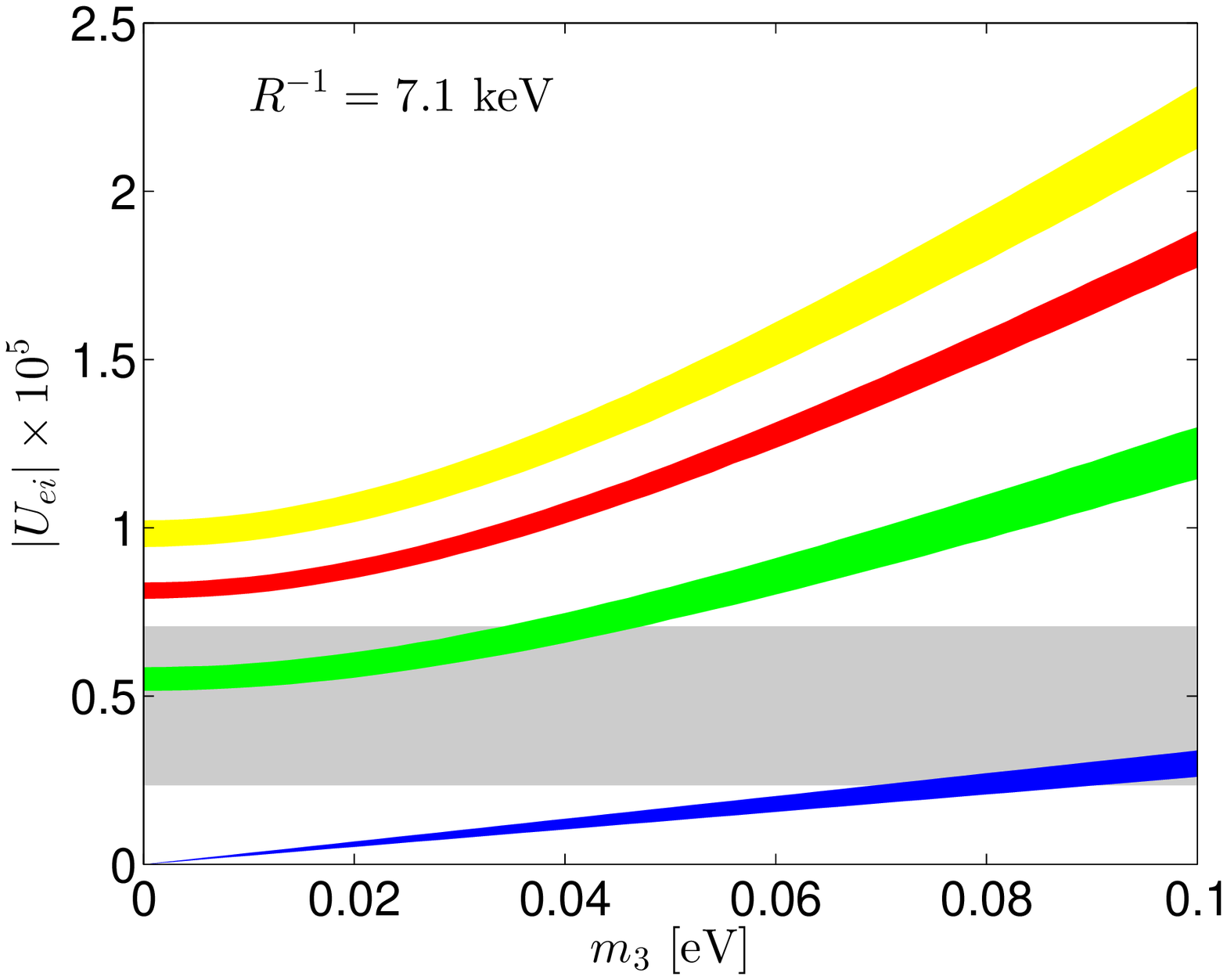}
\caption{\label{fig:U} Allowed ranges (3$\sigma$ C.L.) of active-sterile mixing parameters for the first KK sterile neutrinos ($U_{e4}$ red,
$U_{e5}$ green, $U_{e6}$ blue, and $\theta_{\rm eff}$ yellow). Here normal mass ordering is assumed for the left column, inverted for the right column. For the upper and middle panels, the shaded areas indicate the $2\sigma$ ranges of the active-sterile mixing matrix element from a recent 3+1 global fit~\cite{Kopp:2013vaa}, whereas for the lower panel, the shaded areas correspond to the uncertainty of the mixing angle required for the claimed $7.1~{\rm keV}$ WDM~\cite{Boyarsky:2014jta}. The next set of KK states has mixing suppressed by a factor of 2.}
\end{center}
\end{figure}

We illustrate in Fig.~\ref{fig:U} the allowed ranges (3$\sigma$ C.L.) of the active-sterile mixing $U_{e4}$, $U_{e5}$ and $U_{e6}$ with respect to the lightest neutrino mass. Since sterile neutrino states are nearly degenerate in mass at each KK level, it is useful to define an effective mixing angle as $\theta_{\rm eff} \simeq \sqrt{|U^2_{e4}|+|U^2_{e5}|+|U^2_{e6}|}$ describing the active-sterile oscillations at short-baseline experiments. For light sterile neutrinos, the current global fit range of active-sterile mixing at $2\sigma$ (taken from~\cite{Kopp:2013vaa}) is indicated by shaded areas. As for the keV case, we show the 
preferred mixing range from the claimed signal in X-ray searches \cite{Bulbul:2014sua,Boyarsky:2014jta}.
One can read from the plots that, different to many other sterile neutrino models, our extra dimension model is highly predictive and the only relevant parameters are the extra dimension scale $R$ and the absolute neutrino mass. The required active-sterile mixing from current short-baseline neutrino oscillation anomalies can be well accommodated in our framework. If $R^{-1}$ is larger than $\sim 1~{\rm eV}$, a degenerate mass spectrum of active neutrinos is desired to get sizable mixing. For example, $\theta_{\rm eff} \simeq 0.1 $
is achieved for $m_{\rm lightest} \geq 0.08~{\rm eV} $.

\begin{figure}[t]
\begin{center}\vspace{0.cm}
\includegraphics[width=.45\textwidth]{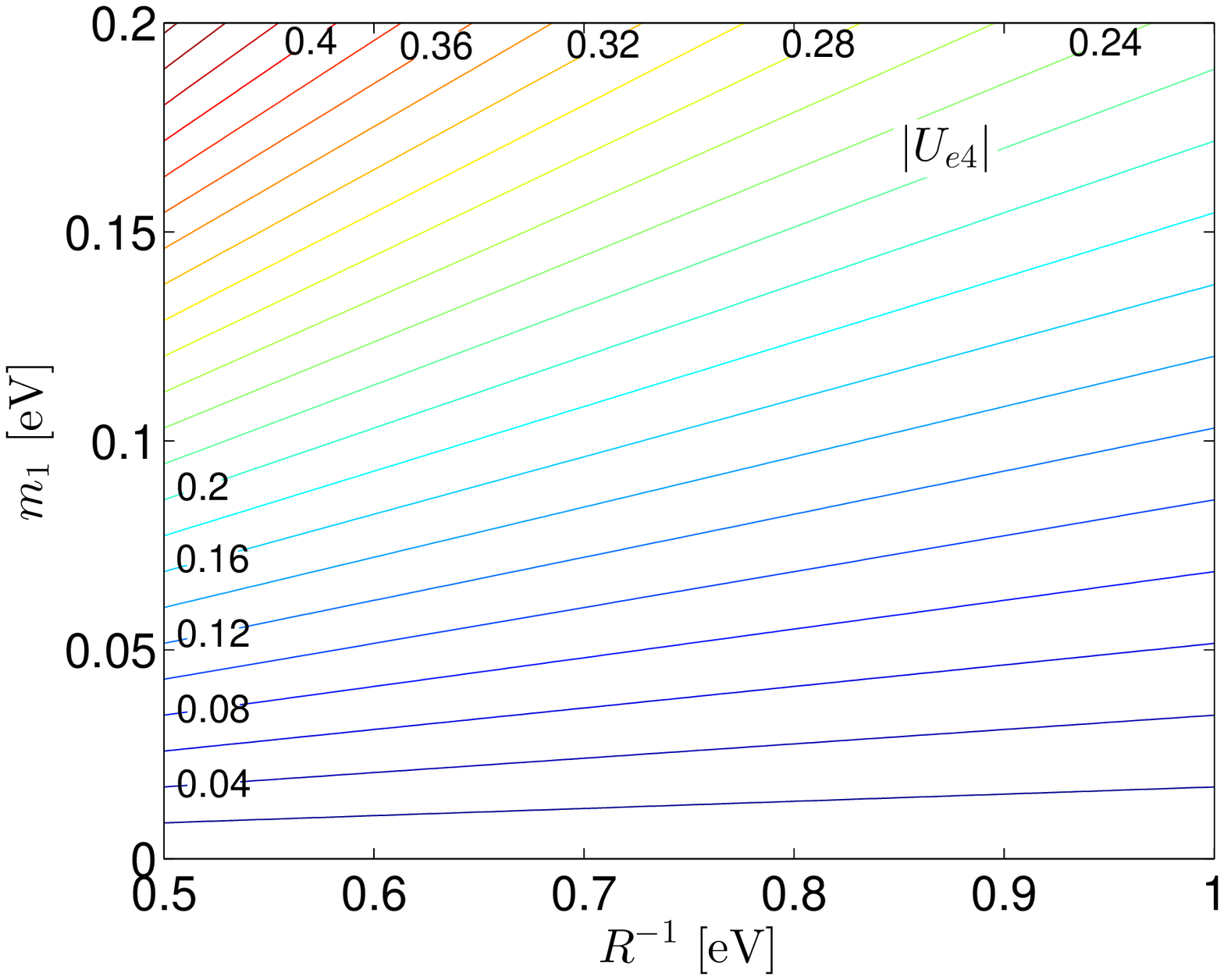}\hspace{5mm}
\includegraphics[width=.45\textwidth]{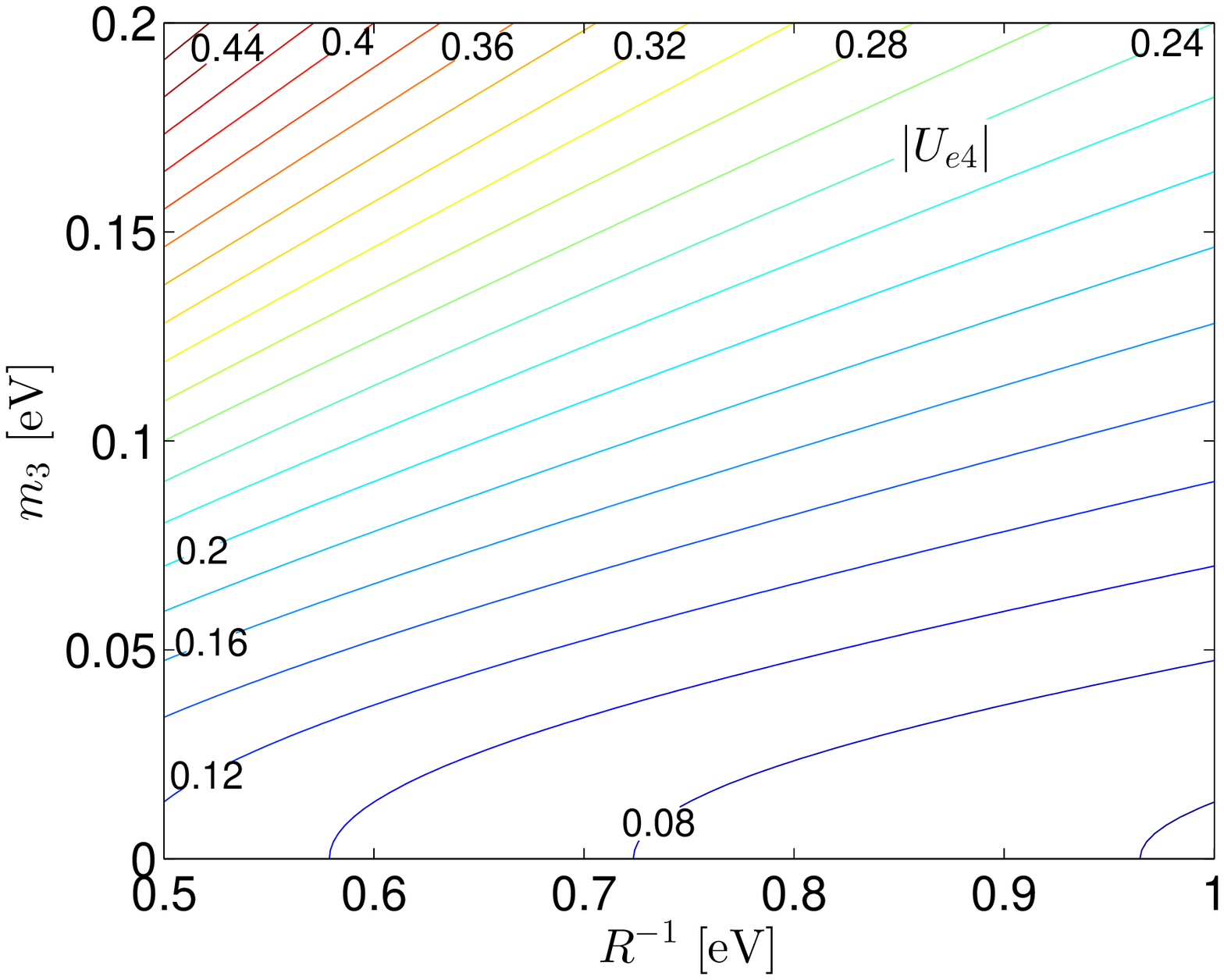}
\includegraphics[width=.45\textwidth]{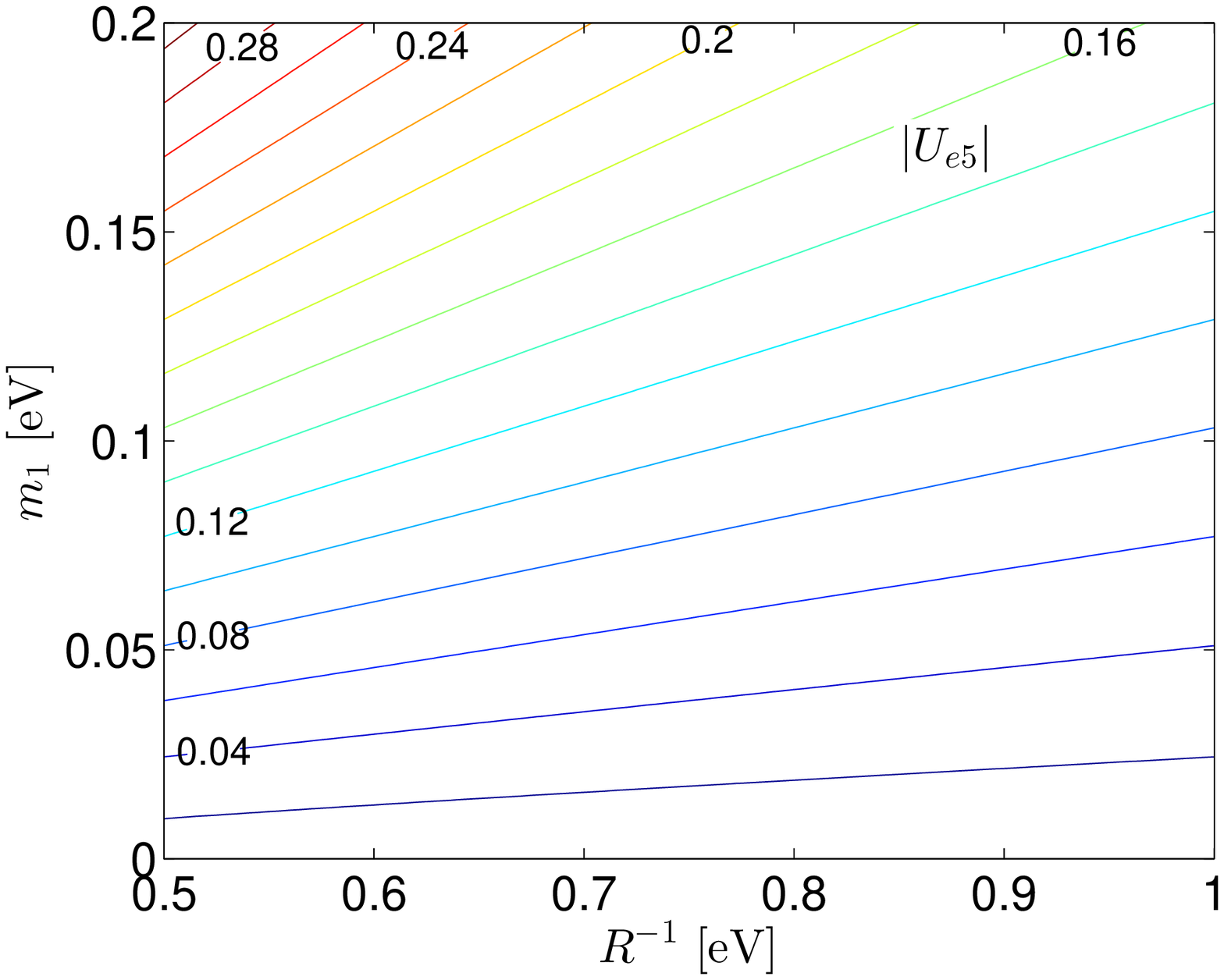}\hspace{5mm}
\includegraphics[width=.45\textwidth]{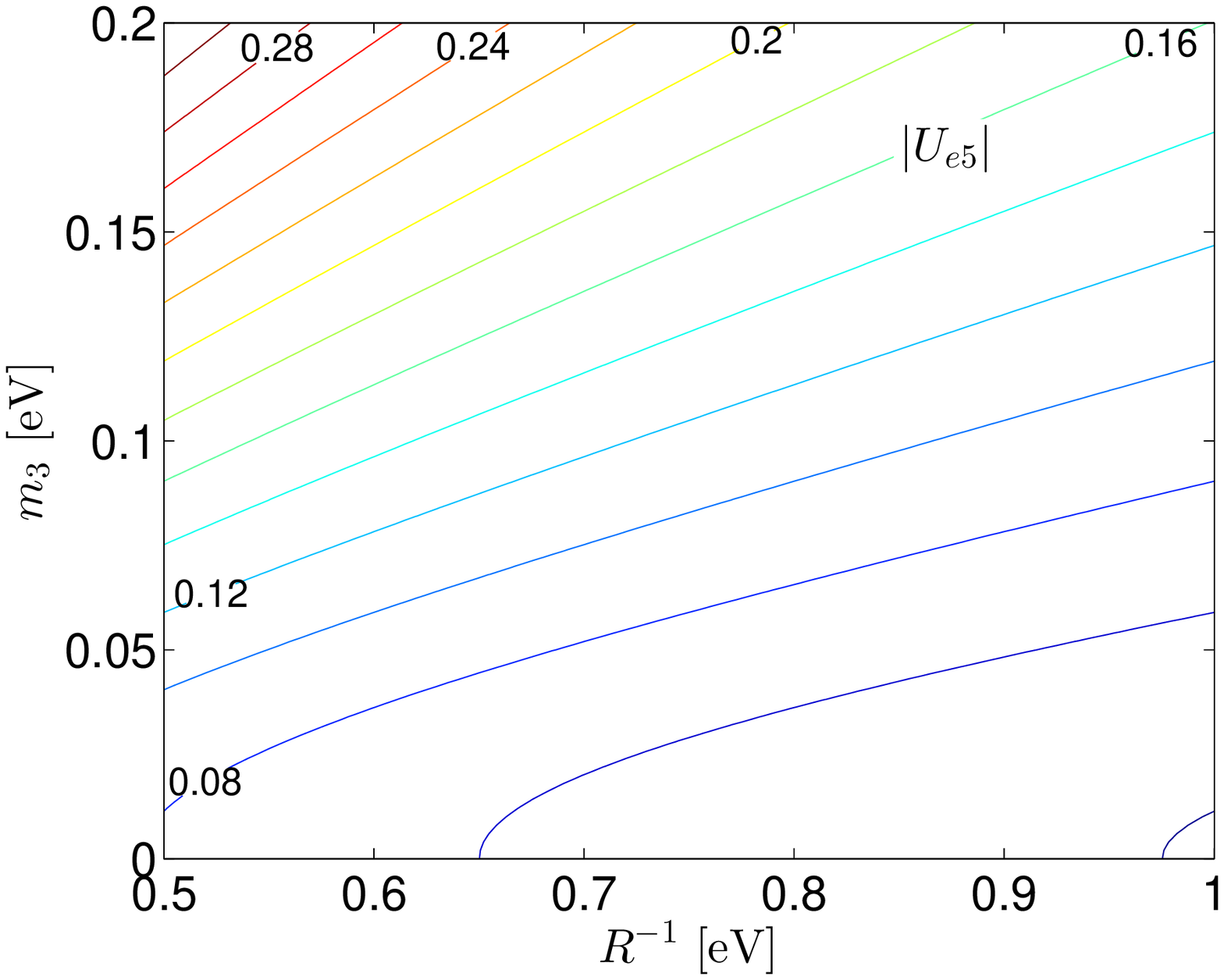}
\includegraphics[width=.45\textwidth]{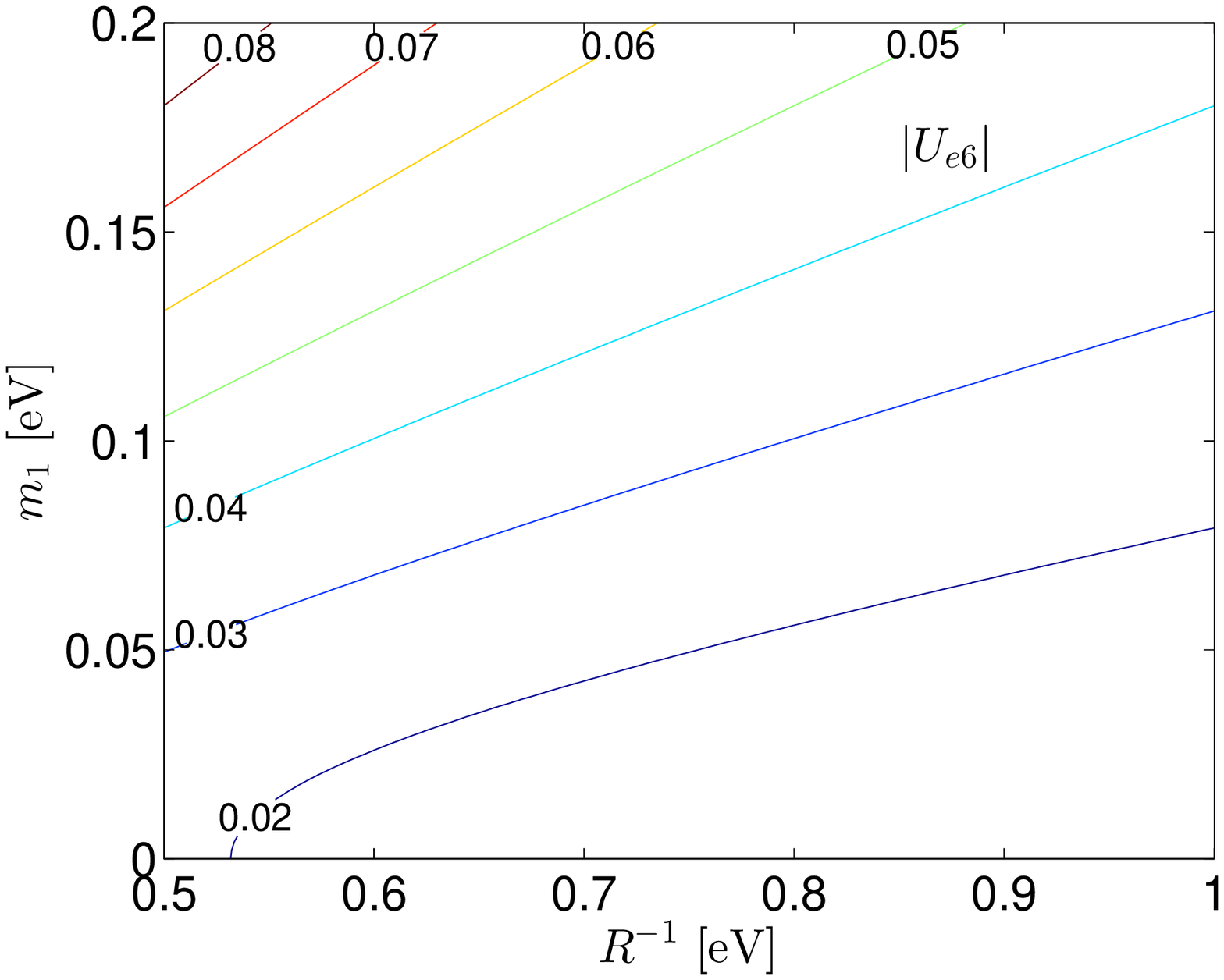}\hspace{5mm}
\includegraphics[width=.45\textwidth]{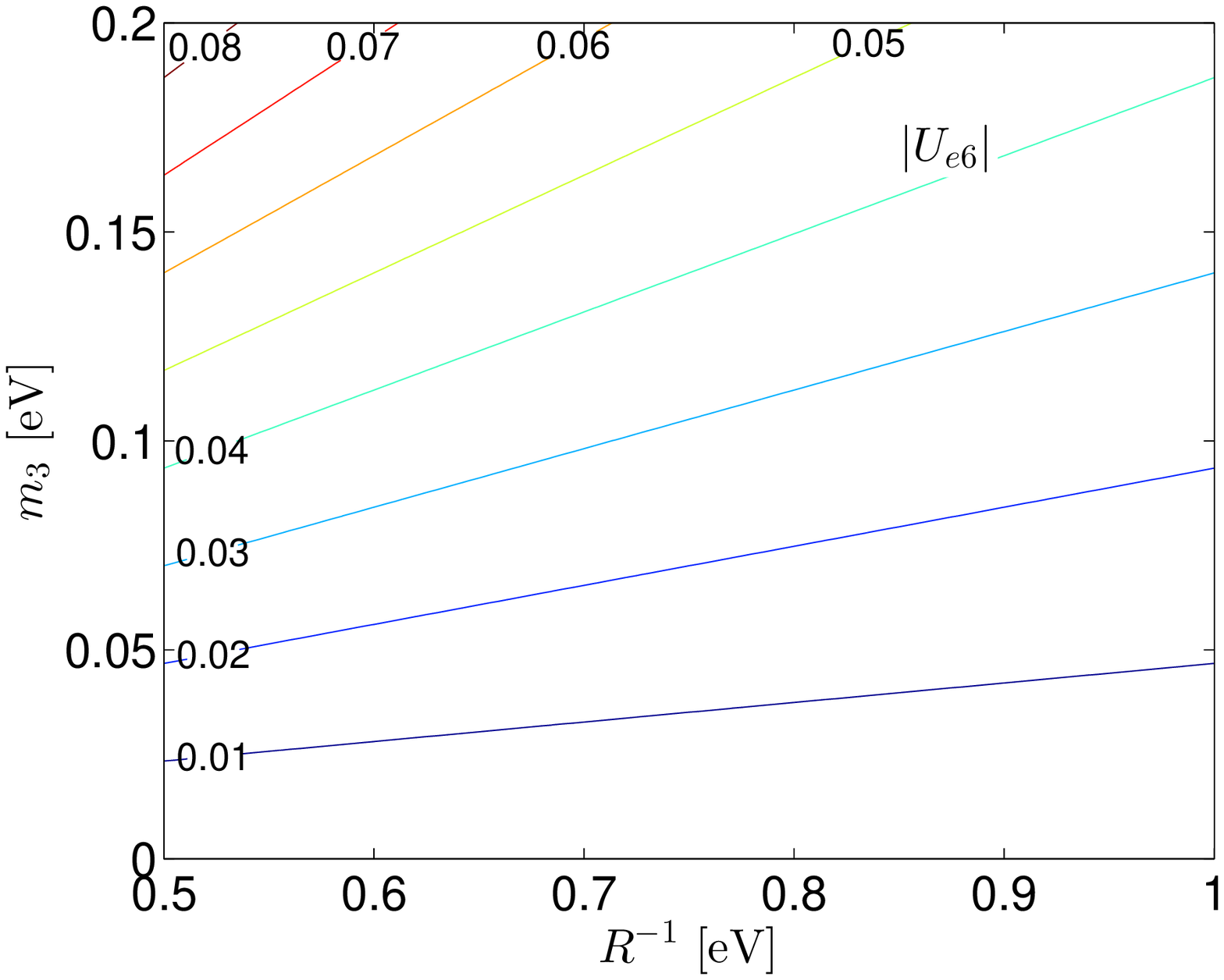}
\caption{\label{fig:UeV} Entries in the first row of $U$ for the normal mass ordering (left column) and inverted mass ordering (right column).}
\end{center}
\end{figure}

\begin{figure}[h]
\begin{center}\vspace{0.cm}
\includegraphics[width=.4\textwidth]{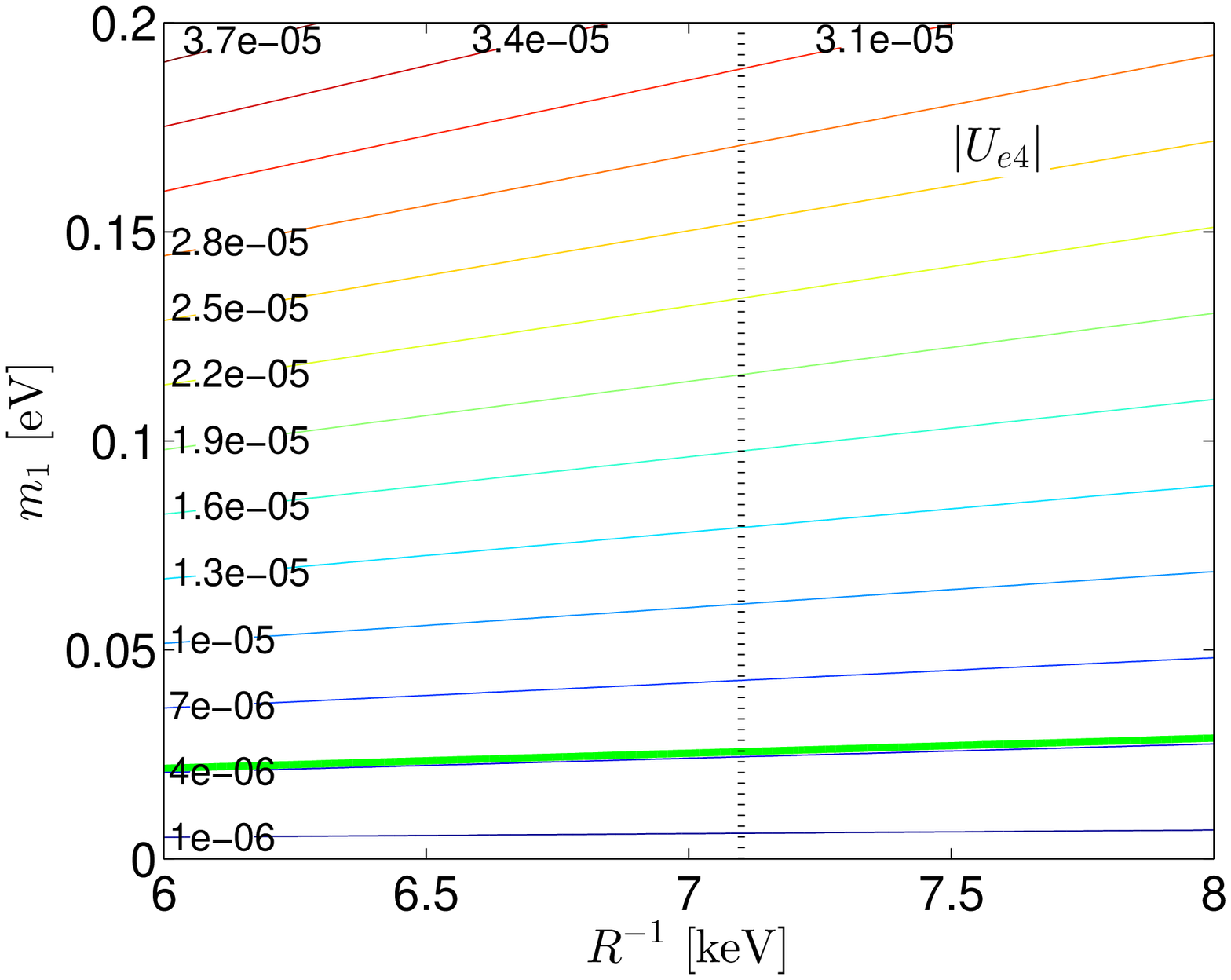}\hspace{5mm}
\includegraphics[width=.4\textwidth]{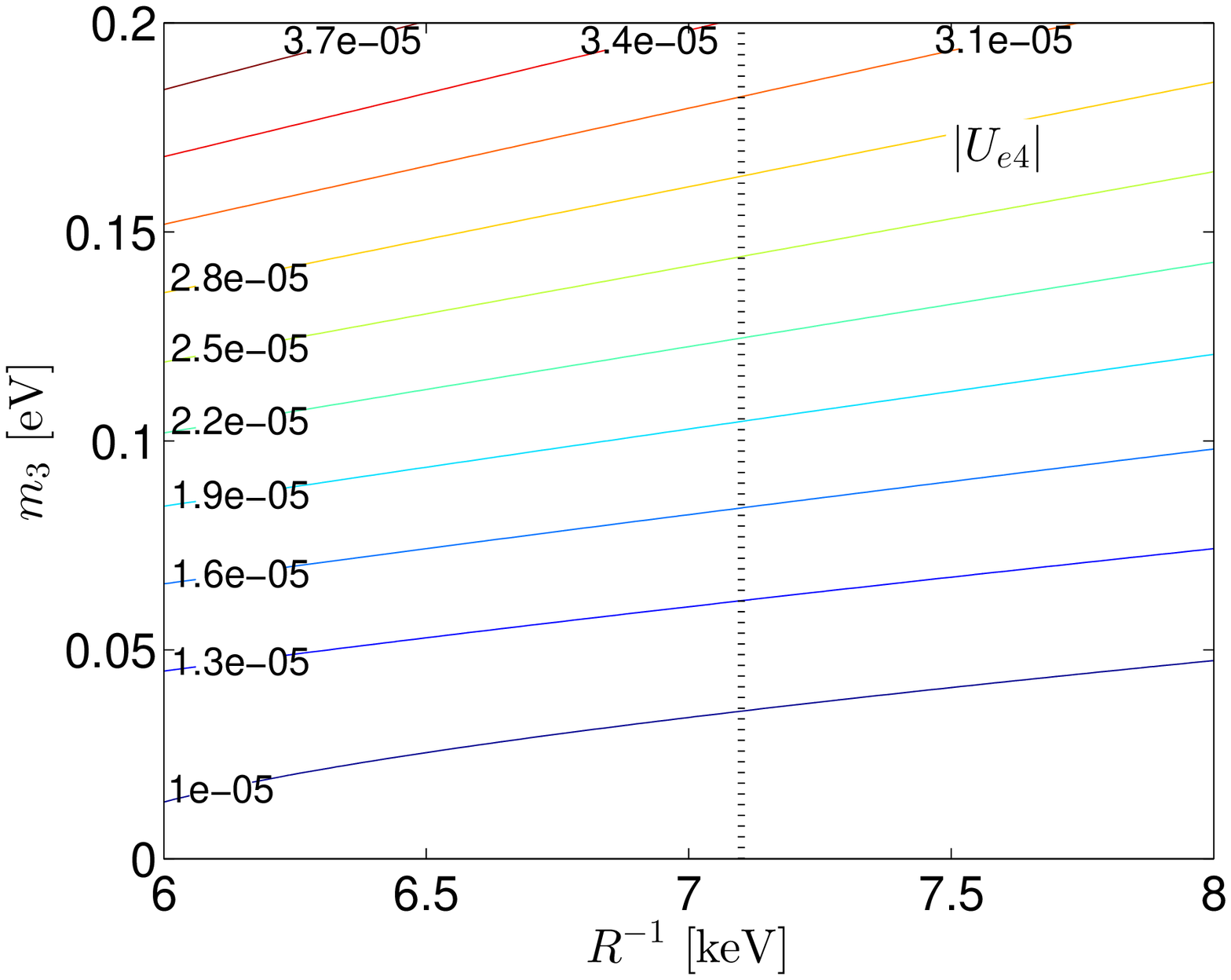}
\includegraphics[width=.4\textwidth]{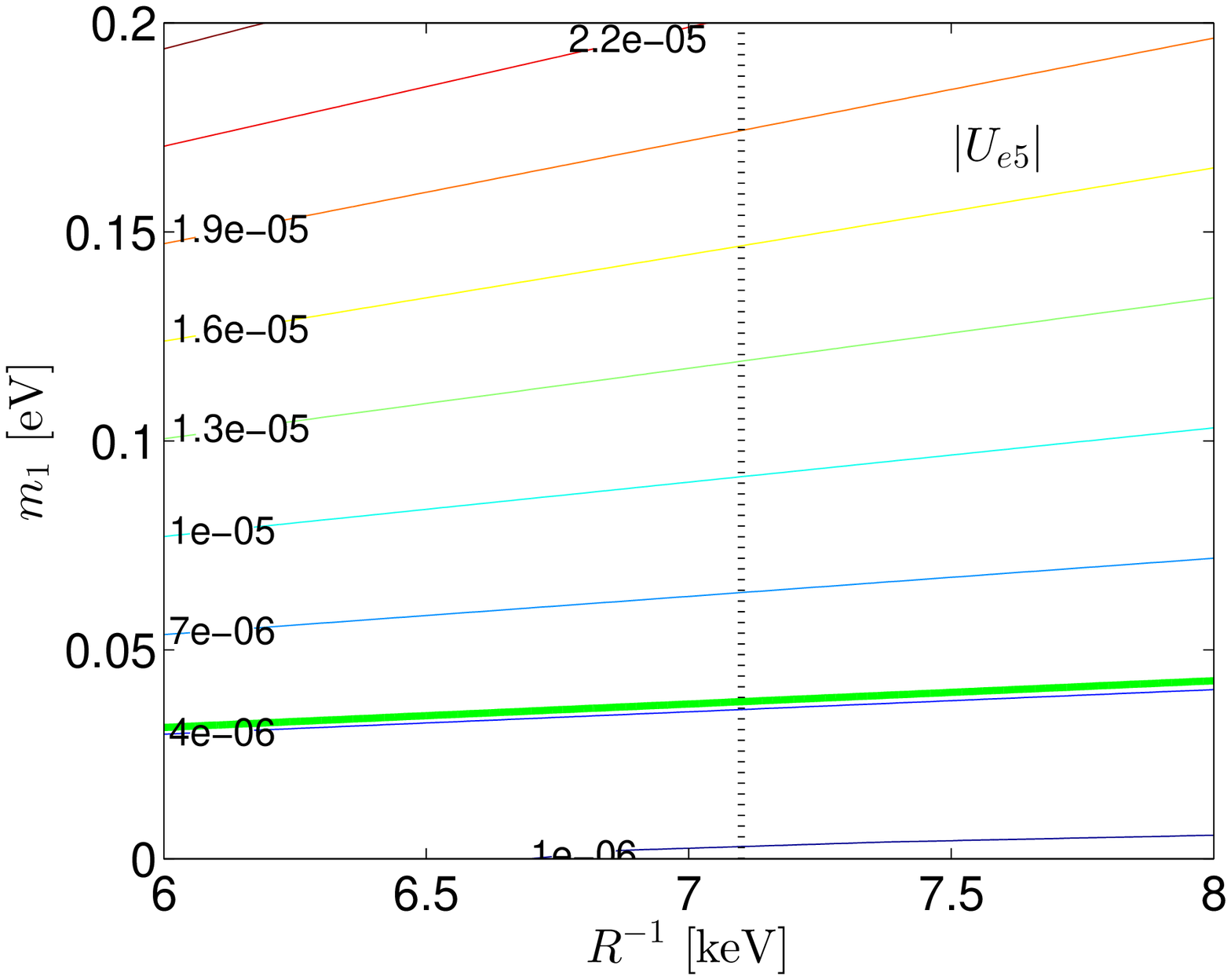}\hspace{5mm}
\includegraphics[width=.4\textwidth]{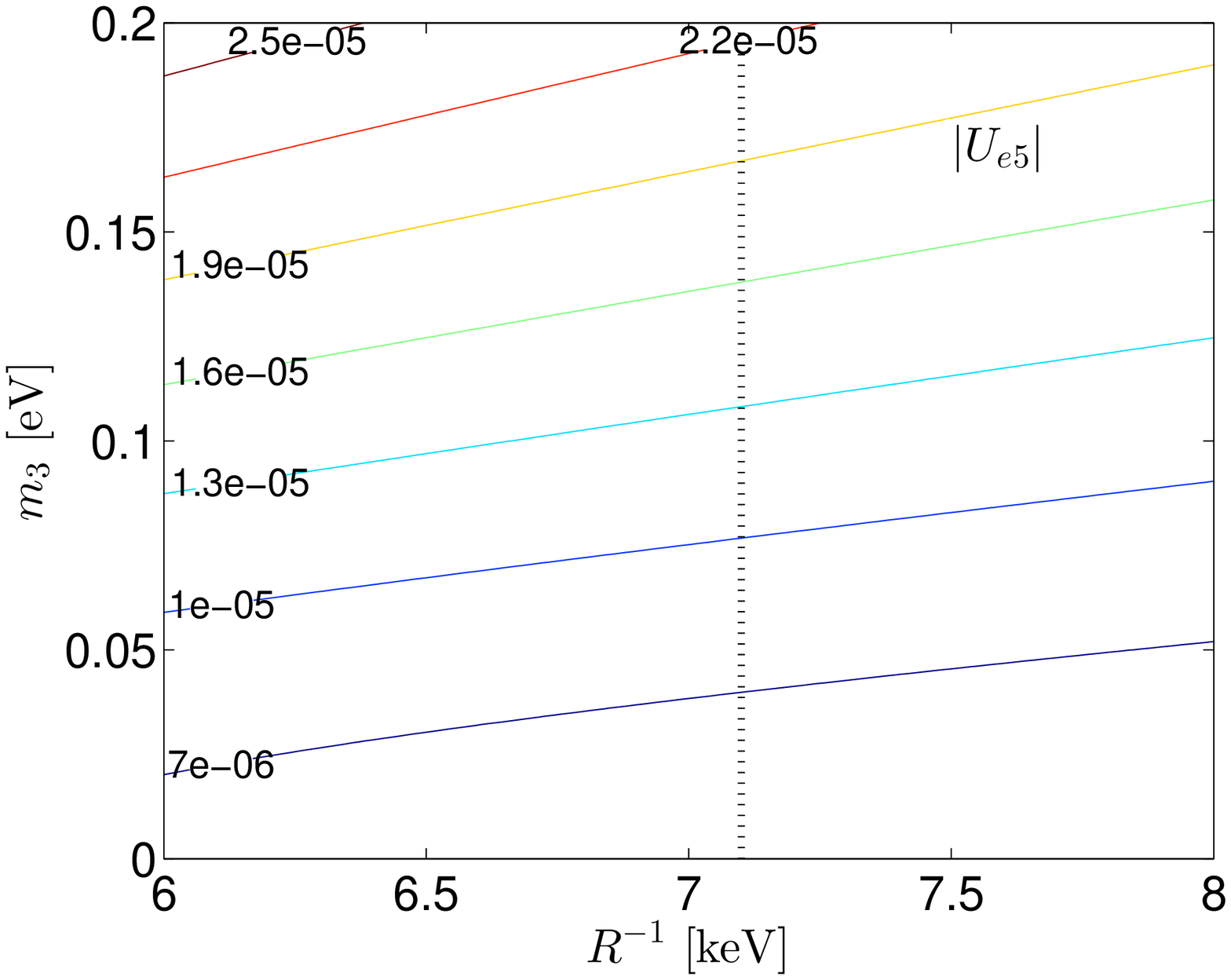}
\includegraphics[width=.4\textwidth]{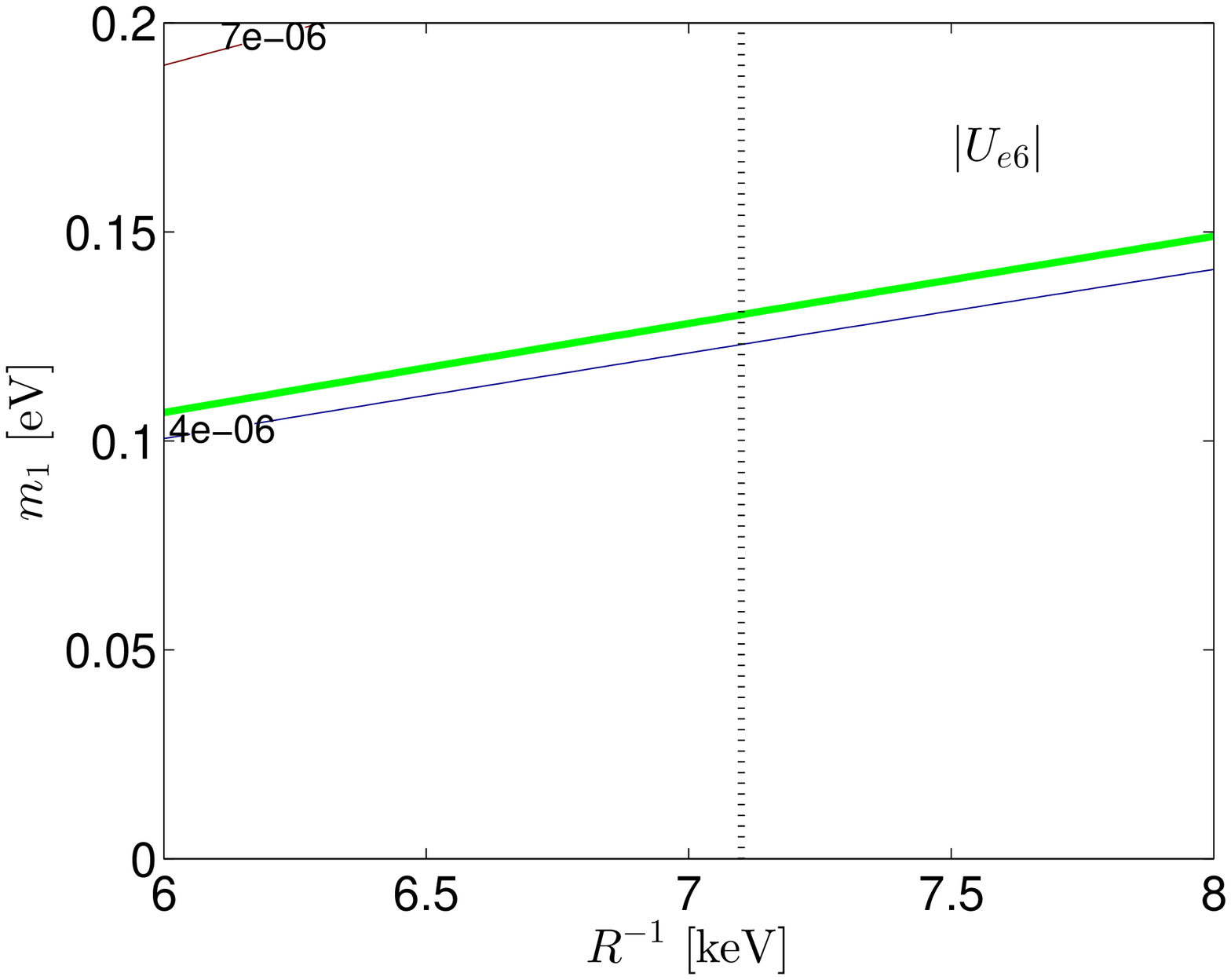}\hspace{5mm}
\includegraphics[width=.4\textwidth]{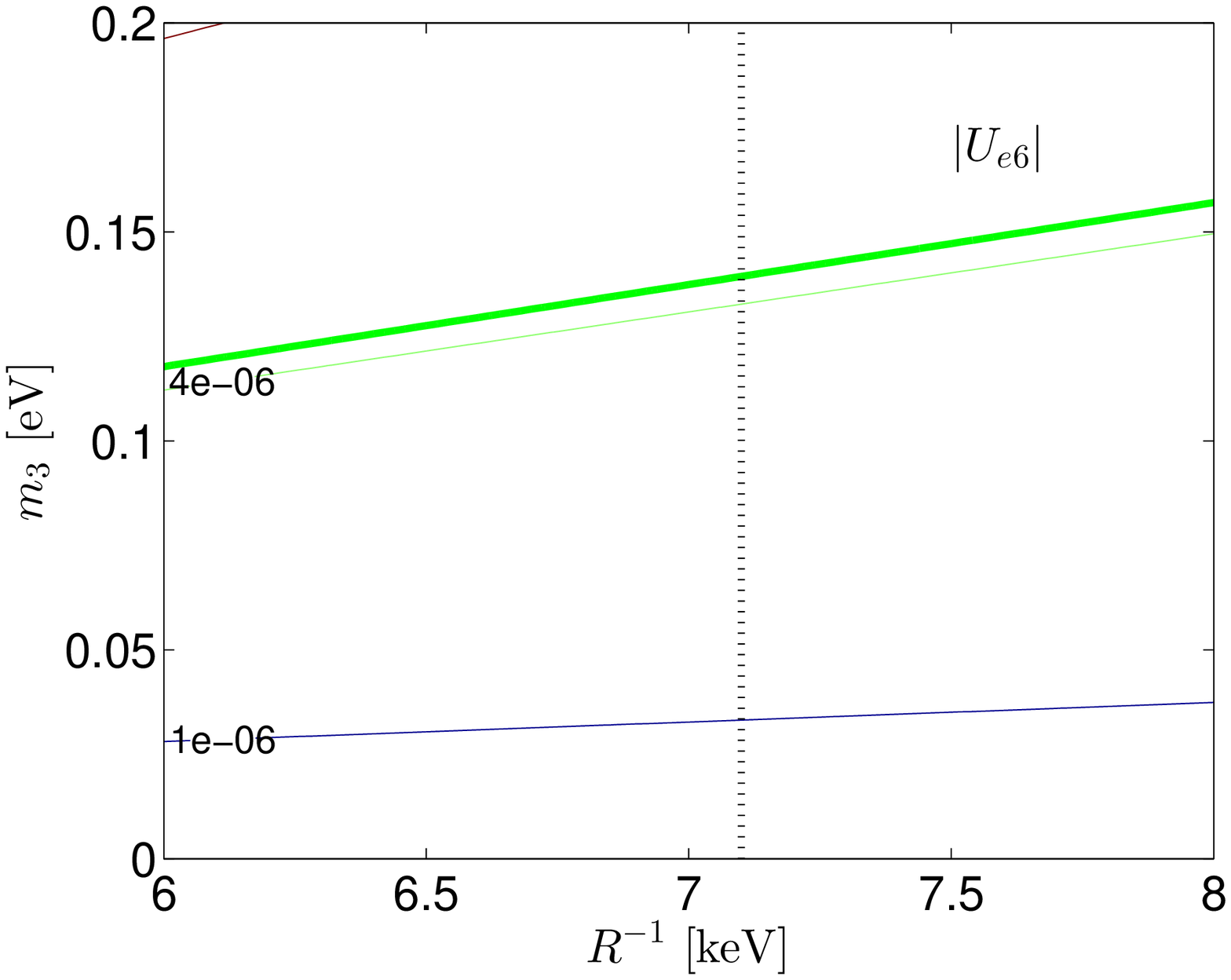}
\caption{\label{fig:UkeV} The entries of $U$ for normal ordering (left column) and inverted ordering (right column). The green thick line corresponds to the mixing angle related to the $7.1~{\rm keV}$ sterile neutrino, while the dotted lines indicate $R^{-1}=7.1~{\rm keV}$. }
\end{center}
\end{figure}

We further show in Figs.~\ref{fig:UeV} and \ref{fig:UkeV} the predicted active-sterile mixing
for the first KK mode in the $R^{-1}$-$m_{1,3}$ plane.
\noindent 
In the plots, all neutrino parameters are fixed to their best-fit values. Similar to Fig.~\ref{fig:U}, the larger the size of extra dimension, the more sizable active-sterile mixing one could expect. Although at the each KK threshold the sterile states are nearly degenerate, their mixing with active neutrinos can be quite different, which is caused by the flavor structure of $m_{\rm D}$. A general feature can be found that $|U_{e4}| > |U_{e5}| > |U_{e6}|$. For keV sterile neutrinos, one observes from Fig.~\ref{fig:UkeV} that
favorable values of $\sim 10^{-6}$ for the mixing matrix elements
can be achieved in the normal hierarchy case for both $U_{e4}$, $U_{e5}$ and $U_{e6}$,
whereas the mixing matrix elements $U_{e4}$ and $U_{e5}$ are too large in the inverted hierarchy case.

\section{phenomenology}
\label{sec:pheno}

The presence of KK sterile neutrinos implies phenomenology in low-energy neutrino experiments.
We proceed to discuss some of the signatures of KK neutrinos in short-baseline oscillation experiments
and in beta spectra as measurable by KATRIN-like experiments.

\subsection{Short-baseline neutrino oscillations}

\begin{figure}[h]
\begin{center}\vspace{0.cm}
\includegraphics[width=.45\textwidth]{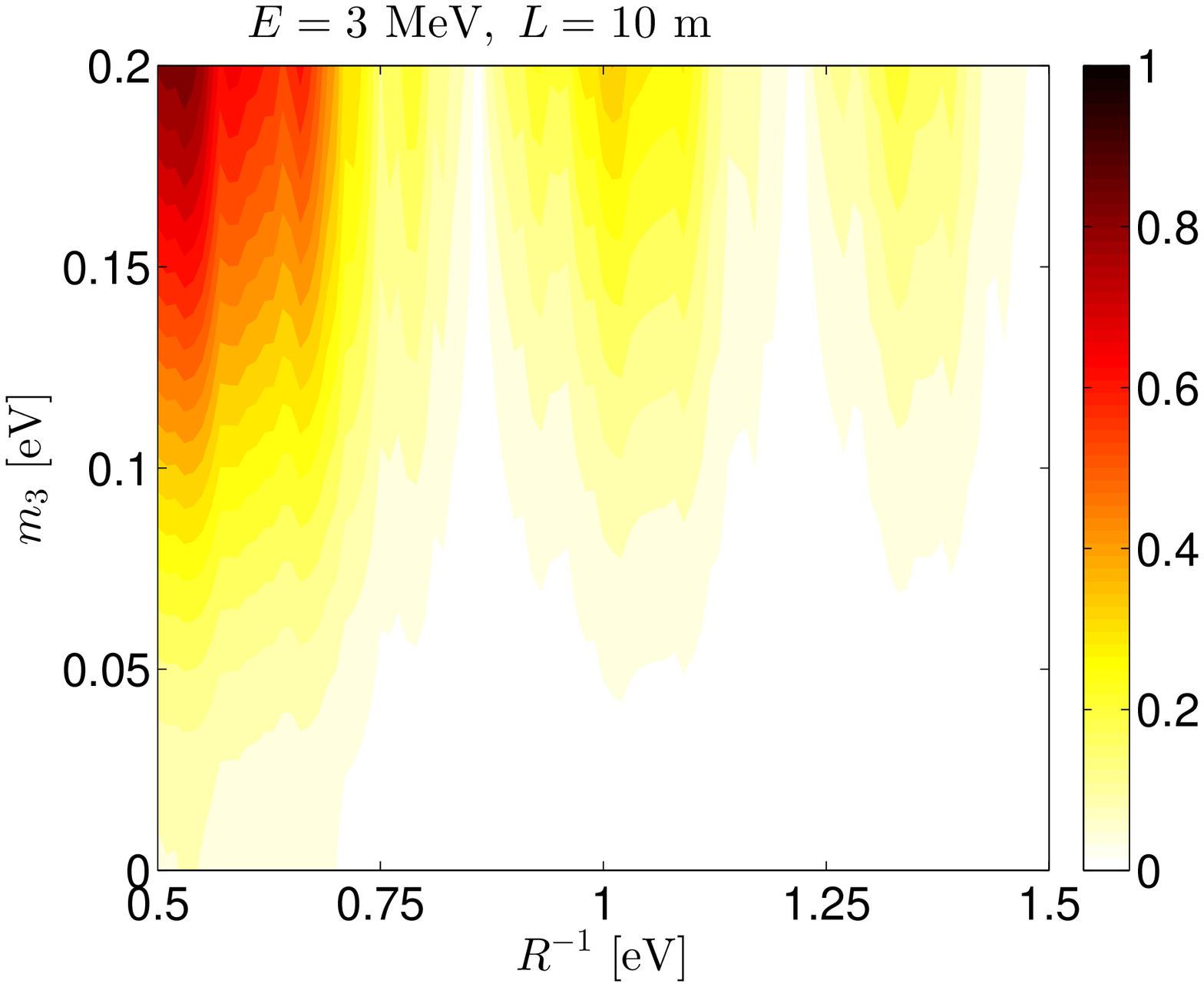}\hspace{5mm}
\includegraphics[width=.45\textwidth]{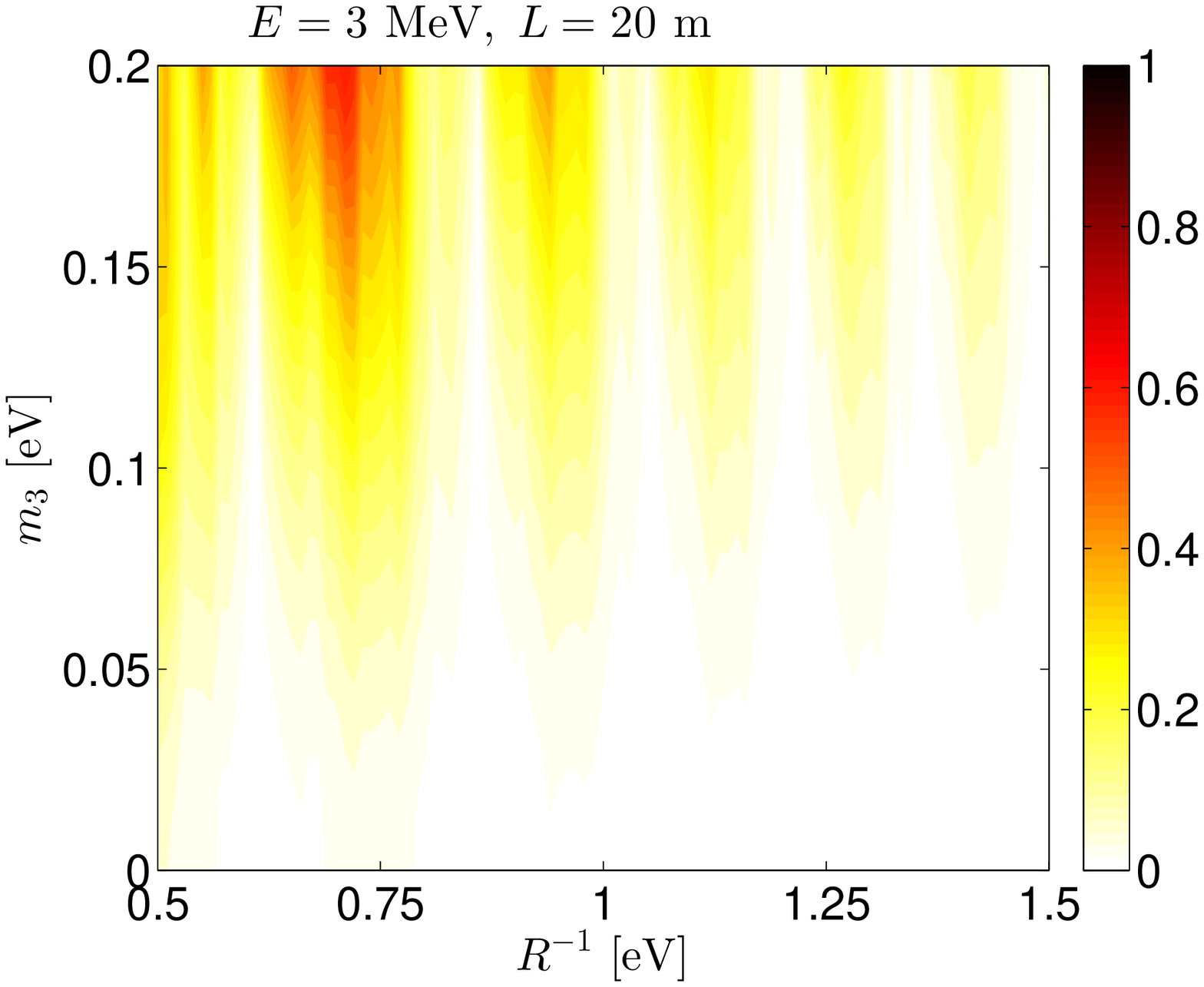}
\includegraphics[width=.45\textwidth]{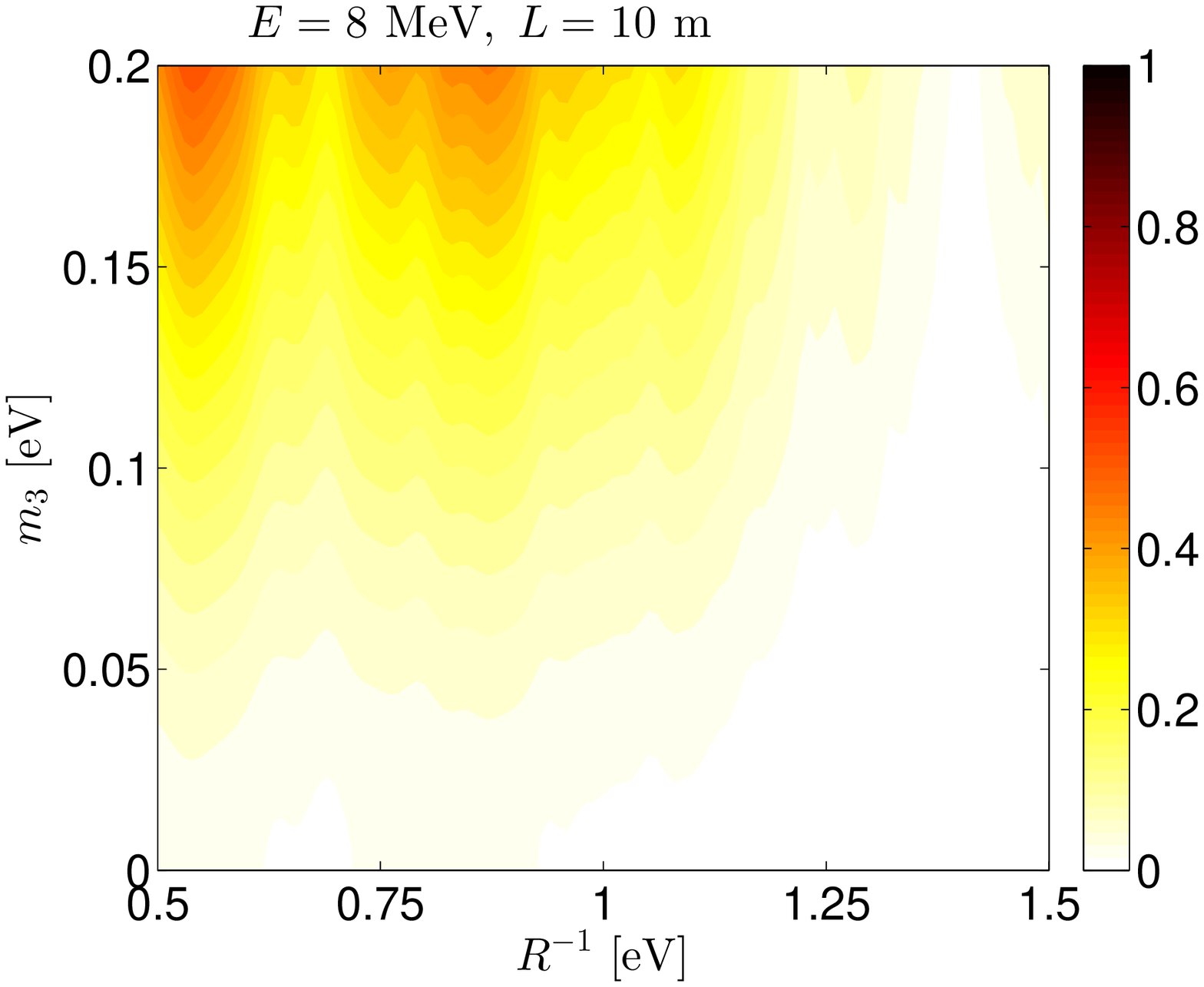}\hspace{5mm}
\includegraphics[width=.45\textwidth]{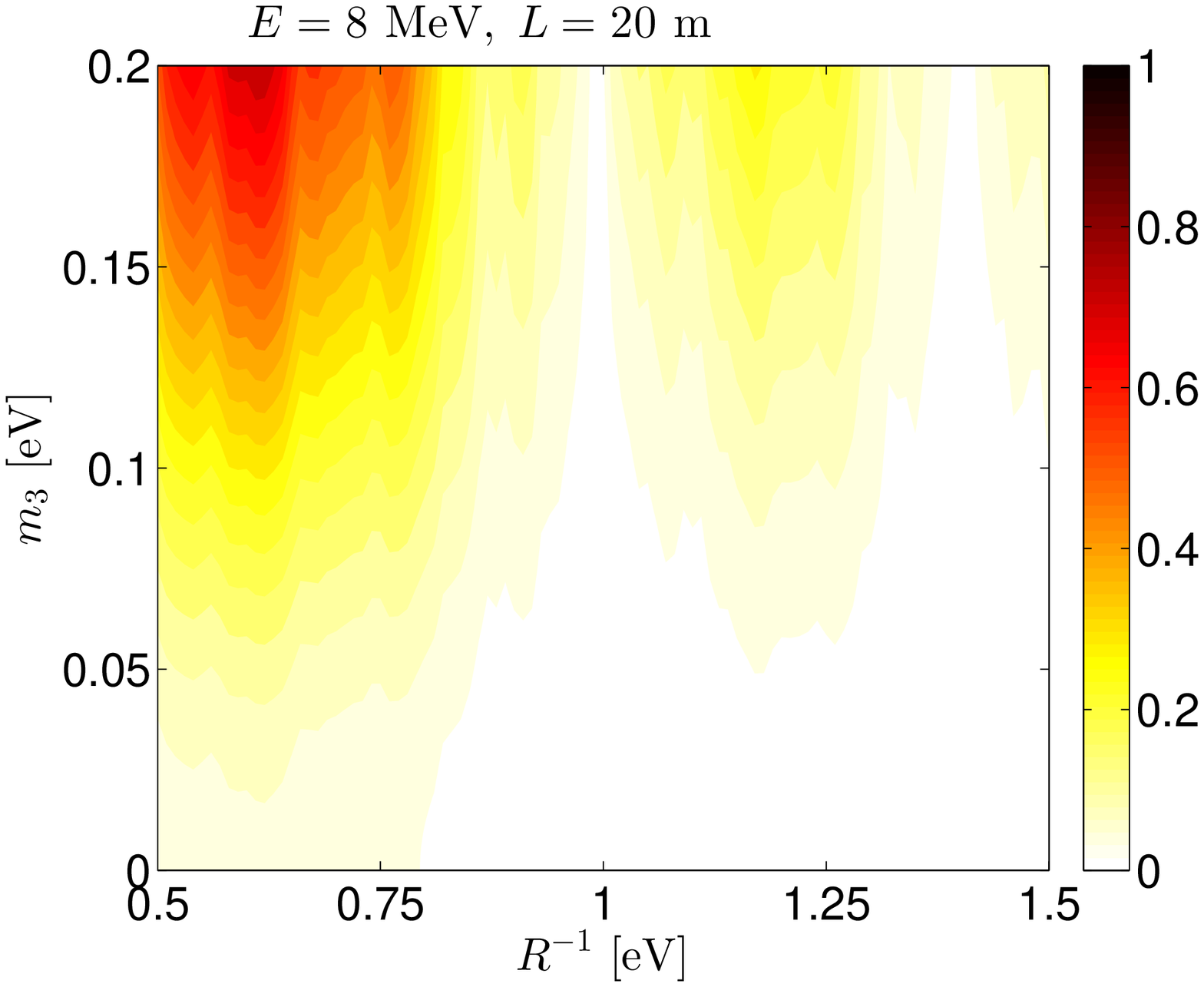}
\caption{\label{fig:PeeIH} The transition probability $1-P_{ee}$ for reactor antineutrinos. Here inverted mass ordering is assumed. The baseline and energy are labeled on each plot.}
\end{center}
\end{figure}

\begin{figure}[h]
\begin{center}\vspace{0.cm}
\includegraphics[width=.45\textwidth]{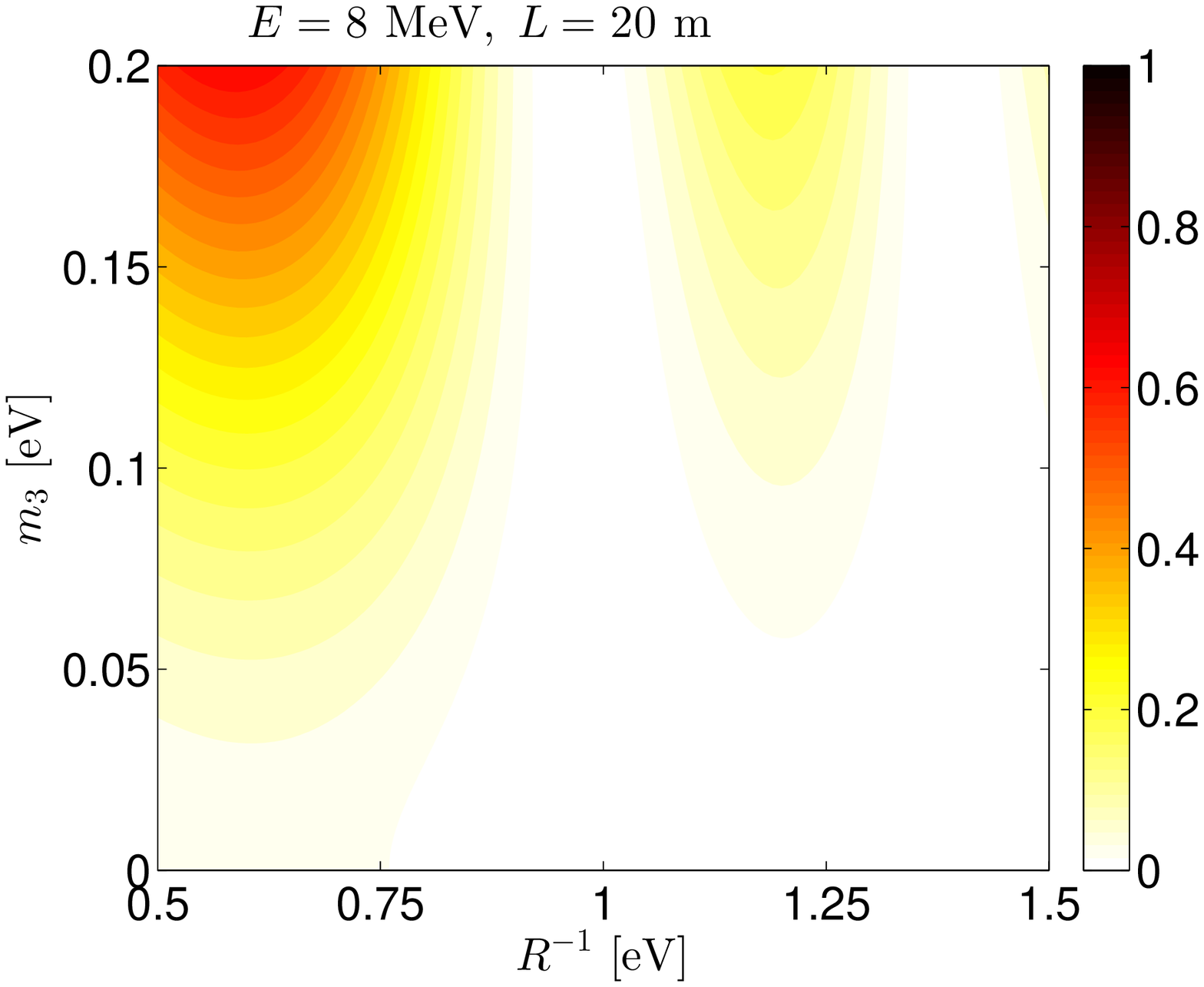}\hspace{5mm}
\includegraphics[width=.45\textwidth]{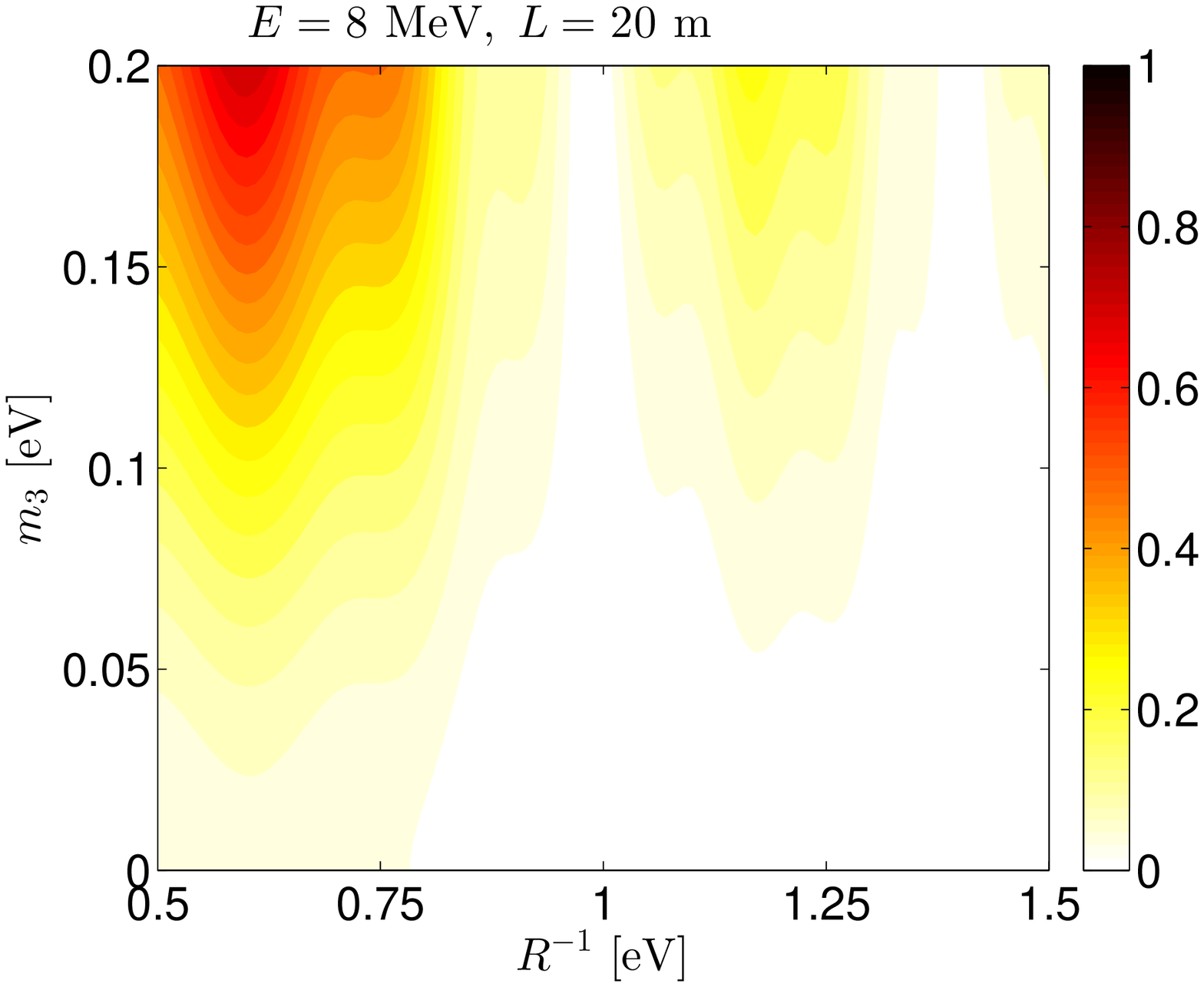}
\includegraphics[width=.45\textwidth]{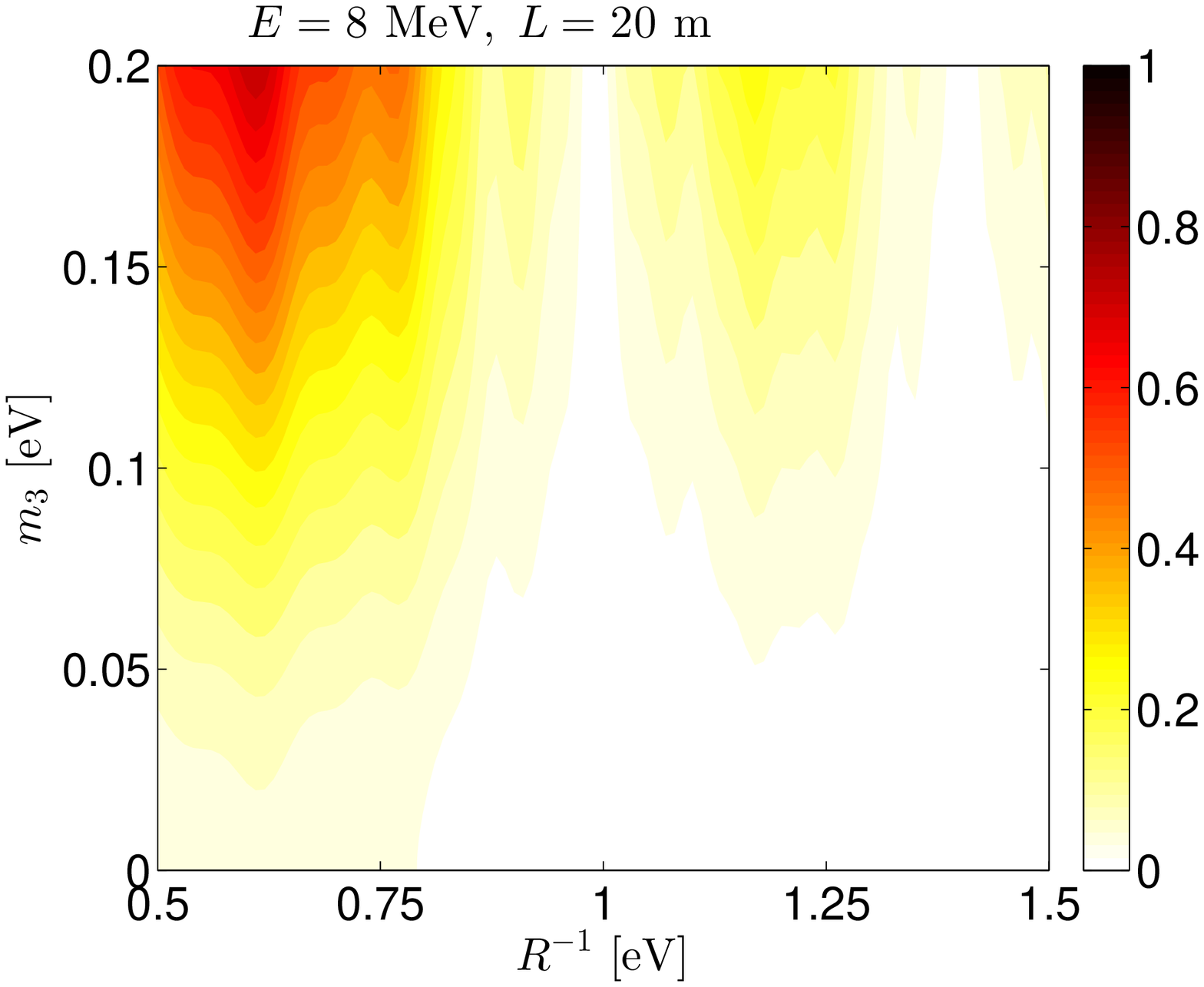}\hspace{5mm}
\includegraphics[width=.45\textwidth]{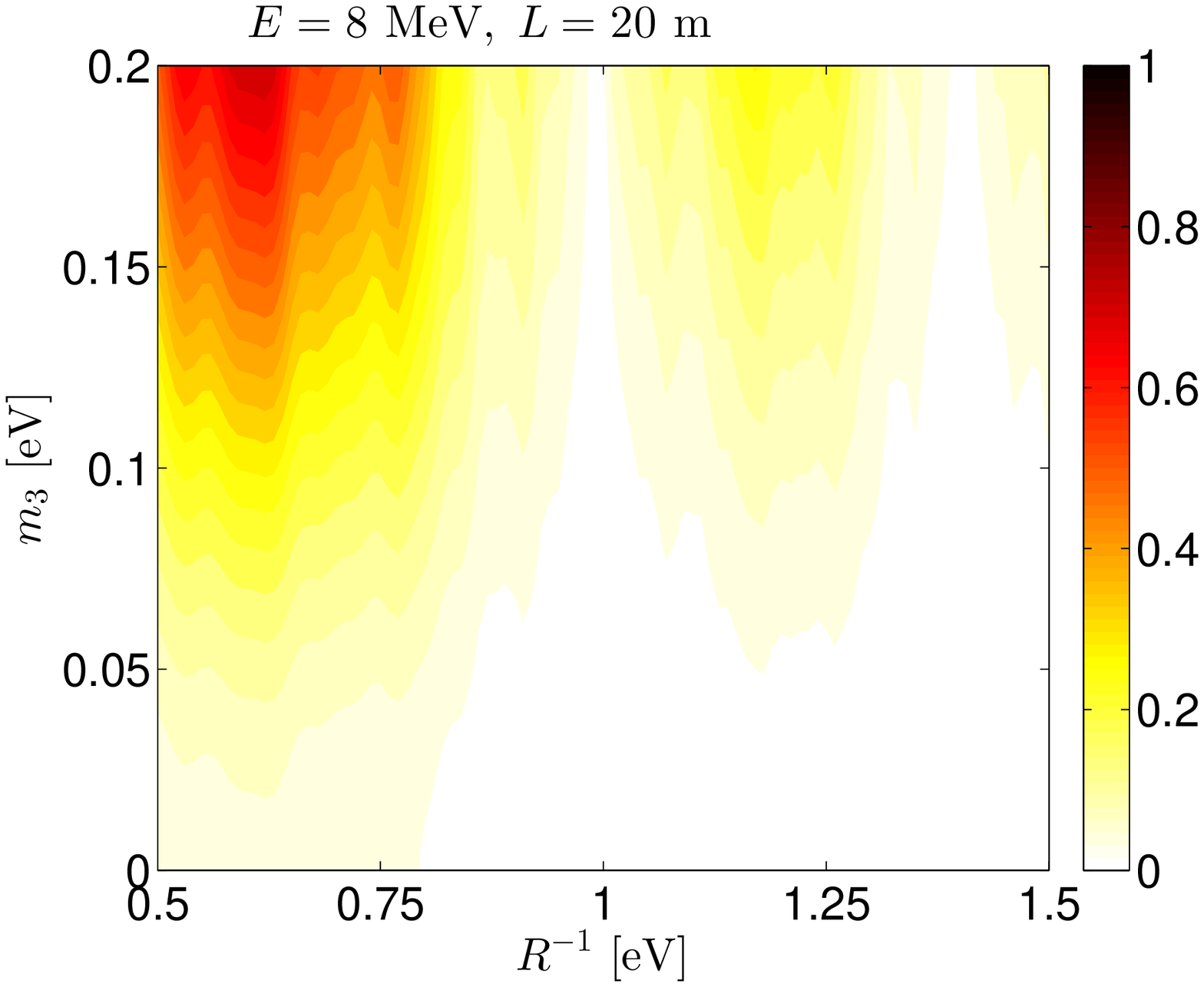}
\caption{\label{fig:PeeIHtest}The lower right plot of Fig.\ \ref{fig:PeeIH} with 1, 2, 3, and 4 KK
modes.}
\end{center}
\end{figure}

Compared to the standard three flavor oscillations in vacuum, the transition amplitude is modified to (see also Ref.\ \cite{Machado:2011jt} for an analysis of neutrino oscillations in a very similar
framework)
\begin{eqnarray}\label{eq:P}
{\cal A}(\nu_\alpha \to \nu_\beta) = \sum^3_{i=1} \left[ V^*_{\alpha i} V_{\beta i} \exp\left({\rm i}\frac{m^2_i L}{2E}\right) + \sum^\infty_{n=1}
K^*_{(n)\alpha i} K_{(n)\beta i} \exp\left({\rm i}\frac{n^2L}{2ER^2}\right) \right]  ,
\end{eqnarray}
where $E$ is the neutrino energy and $L$ denotes the baseline. 
Squaring the amplitude ${\cal A}(\nu_e \to \nu_e)$ gives the survival
probability $P_{ee}$. 
Since a short-baseline neutrino experiment provides the most promising
signal for light sterile neutrinos, we show in Fig.~\ref{fig:PeeIH}
the transition probability $1-P_{ee}$ for reactor antineutrinos at different baselines.
As expected, the deficit of reactor neutrinos at short distance
\cite{Mention:2011rk,Huber:2011wv} 
can be easily explained due to the oscillation of electron neutrinos
into KK sterile neutrinos\footnote{We refer here of course to the
  so-called reactor anomaly: a re-evaluation of reactor neutrino
  fluxes resulted in an averaged ca.\ 3\, \% reduction of past experiments,
  which can be easily interpreted \cite{Kopp:2013vaa} in terms of oscillation into sterile
neutrinos with mass scale eV and mixing of order 0.1.}.
We truncate in our numerical analysis the KK tower after 5 modes.
To see the effect of the individual KK states, we show in Fig.\ \ref{fig:PeeIHtest} the
transition probability $1 - P_{ee}$ when 1, 2, 3 and 4 KK modes are included.
In practice, only the first few KK modes play a role in neutrino oscillations and effects from
heavier KK modes (e.g.\ for $n>4$) are negligibly small. 
We note that Ref.\ \cite{Girardi:2014gna} has, within a similar extra dimension 
framework, recently obtained limits from reactor neutrino experiments 
on $R$ of around eV. 

\subsection{$\beta$-decay spectrum}

The extra KK states also affect the electron spectrum in beta decays. Let us consider
tritium decay, ${}^3{\rm H}\rightarrow {}^3{\rm He}+e^-+\overline{\nu_e}$, for which the
electron energy spectrum is given by
\begin{eqnarray}\label{eq:dNdE}
\frac{{\rm d}N}{{\rm d}E} & = & F(E)\sum^3_{i=1}
\left|V_{ei}\right|^2(E_0-E)\left[(E_0-E)^2-m^2_i\right]^{\frac{1}{2}}
\Theta(E_0-E-m_i) \nonumber \\
&+&  F(E)\sum^\infty_{n=1}\sum^3_{i=1}
\left|K_{(n)1i}\right|^2(E_0-E)\left[(E_0-E)^2-n^2/R^2\right]^{\frac{1}{2}}
\Theta(E_0-E-n/R)\,,
\end{eqnarray}
where the first line corresponds to the active neutrino contribution
while the second line denotes the contribution from KK sterile
neutrinos. Here $E_0$ is the total decay energy, i.e.\ $E_0 =
18.59~{\rm keV}$, and $F(E)$ is a mass independent function given by
\begin{eqnarray}\label{eq:f}
F(E)=G^2_F \frac{m^5_e}{2\pi^3} \cos^2\theta_C |M|^2 R(Z,E)pE\,.
\end{eqnarray}
Here $M$ is the matrix element and $R(Z,E)$ takes Coulomb interactions 
into account, see \cite{Osipowicz:2001sq}. We can define the Kurie function
\begin{eqnarray}\label{eq:KT}
K(E)=\sqrt{\frac{{\rm d}N/{\rm d}E}{F(E)}}\,,
\end{eqnarray}
which, in the absence of neutrino masses, is a linear function close to
the end-point. We focus here on the potential of the KATRIN experiment \cite{Osipowicz:2001sq}.
The effect of eV-scale sterile neutrinos in KATRIN has been studied in
\cite{Riis:2010zm,Formaggio:2011jg,Esmaili:2012vg}. A modified setup of KATRIN is currently seriously under consideration \cite{SM}, which could access whole beta spectrum and hence be sensitive to
keV-scale neutrinos, with mixing potentially going down to $10^{-4}$. Analyses of the spectrum for
keV neutrinos can be found in \cite{deVega:2011xh,Barry:2014ika}.
We should note that in standard WDM scenarios the mixing is probably too small
to be observable by KATRIN\footnote{This can be evaded if the keV neutrinos have additional
interactions, such as within left-right symmetric theories, see \cite{Barry:2014ika}.}, see
Eq.\ (\ref{eq:WDM}).
However, given the unknown mechanism of warm dark matter generation, or the nature of dark matter in general, the study of keV-scale neutrinos in beta decays
is of course interesting in its own right. We note that 
Ref.\ \cite{BastoGonzalez:2012me} has studied the beta spectrum in a very similar 
extra-dimensional setup, focusing on the low energy part of the spectrum 
and on constraints on $R$.

\begin{figure}[t]
\begin{center}\vspace{0.cm}
\includegraphics[width=1\textwidth]{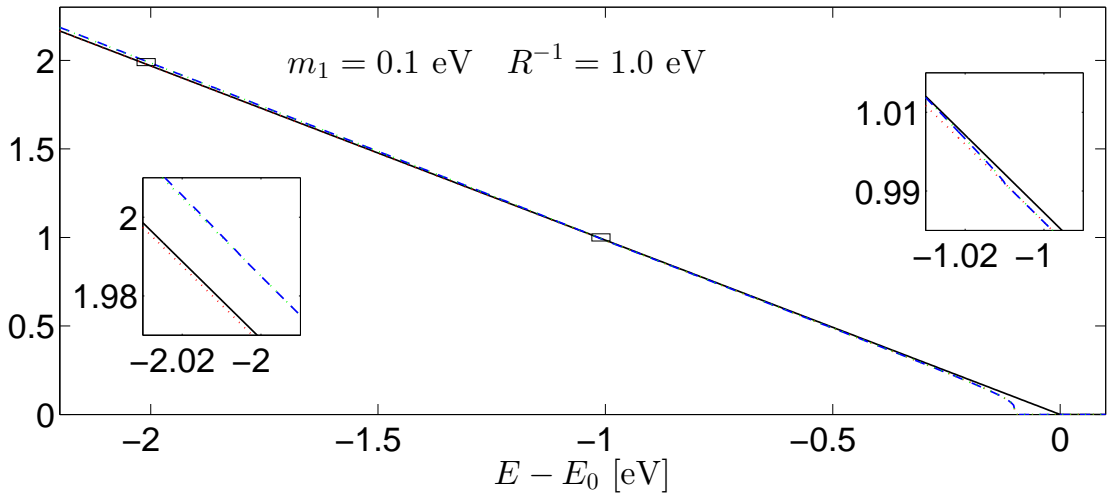}
\caption{\label{fig:Kurie} Kurie function close to end-point of the $\beta$-decay energy spectrum. We have assumed the normal mass ordering together with $m_1=0.1~{\rm eV}$ and $R^{-1}=1~{\rm eV}$. Here the Kurie function for massless (i.e.\ no neutrino masses and no extra dimensions) and massive neutrinos (i.e.\ neutrino mass but no extra dimension) are shown by black solid lines and red dotted lines. The blue dashed lines indicate the case of our extra dimensional model. In order to see the (tiny) effect of the second KK mode, we also show in green dotted line the case with only the lowest KK sterile neutrino switched on.}
\end{center}
\end{figure}

\begin{figure}[t]
\begin{center}\vspace{0.cm}
\includegraphics[width=.49\textwidth]{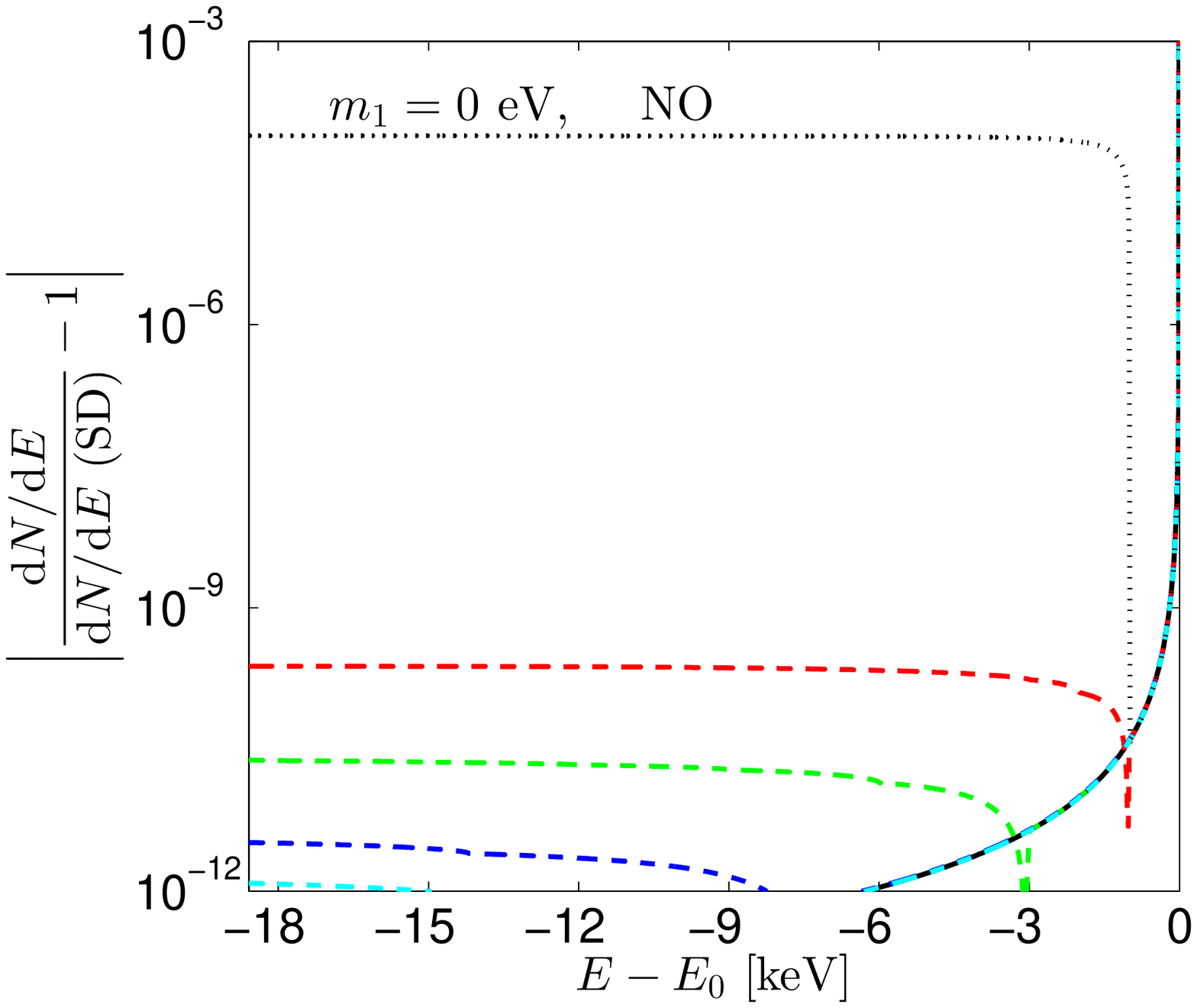}
\includegraphics[width=.49\textwidth]{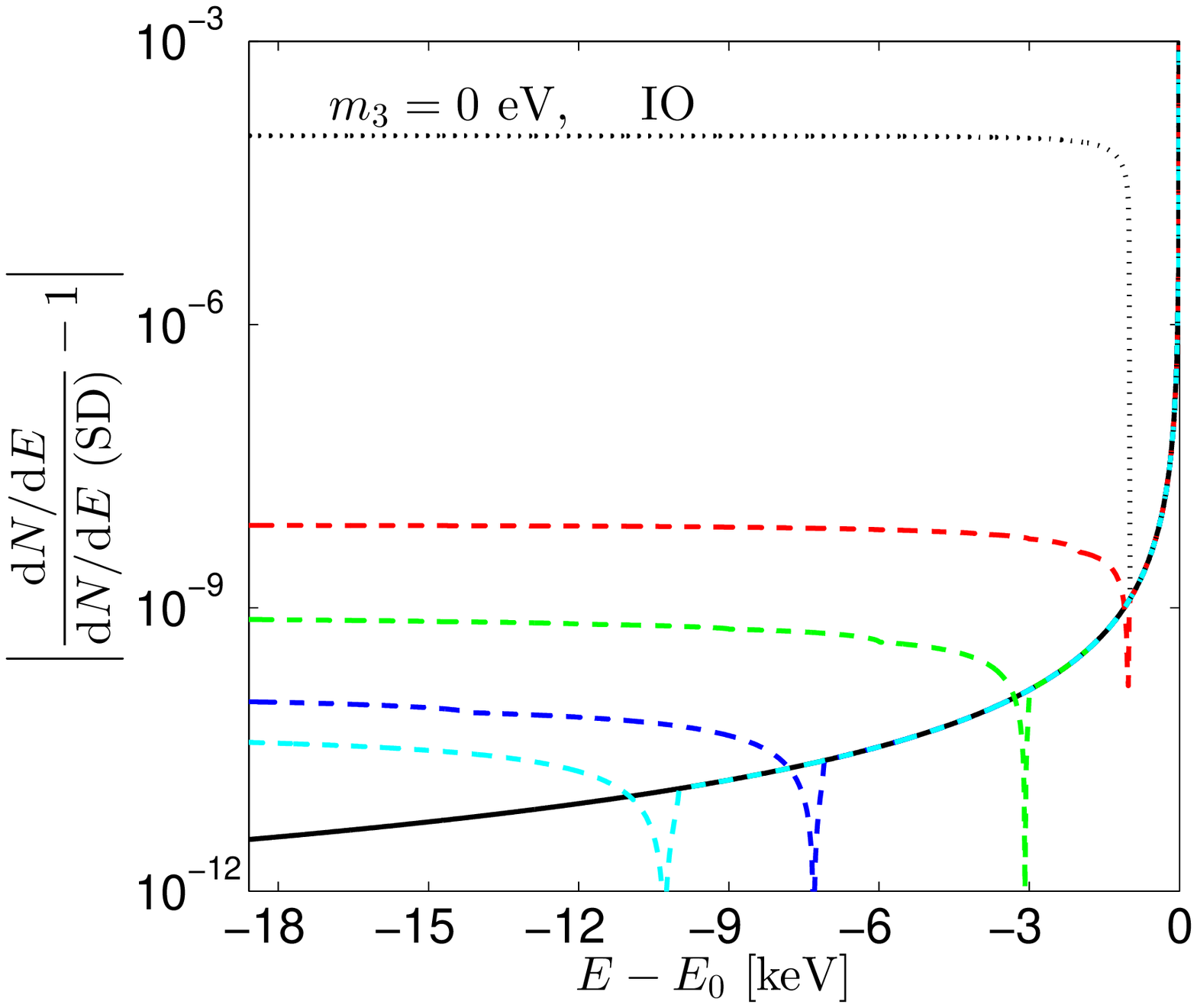}
\includegraphics[width=.49\textwidth]{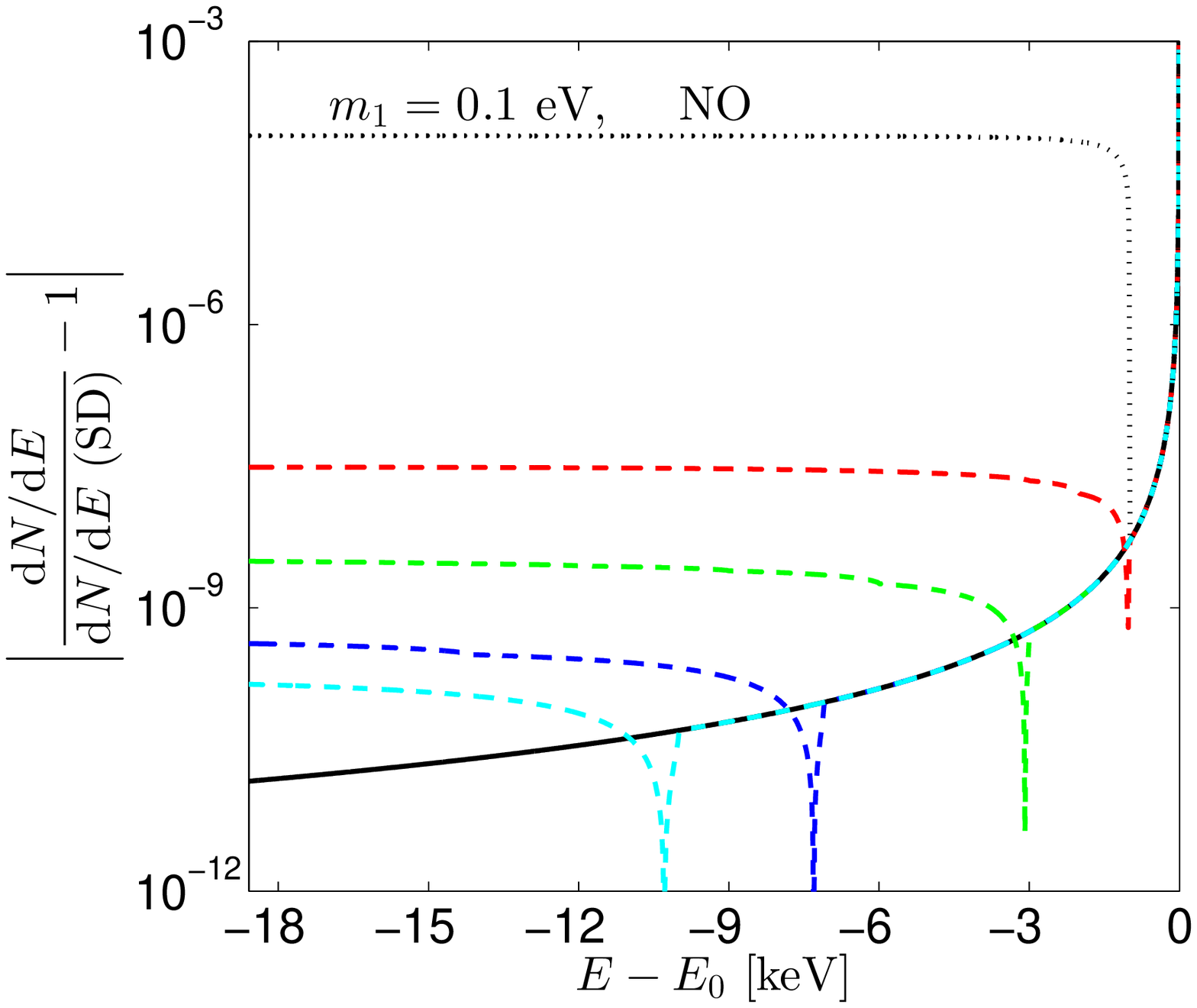}
\includegraphics[width=.49\textwidth]{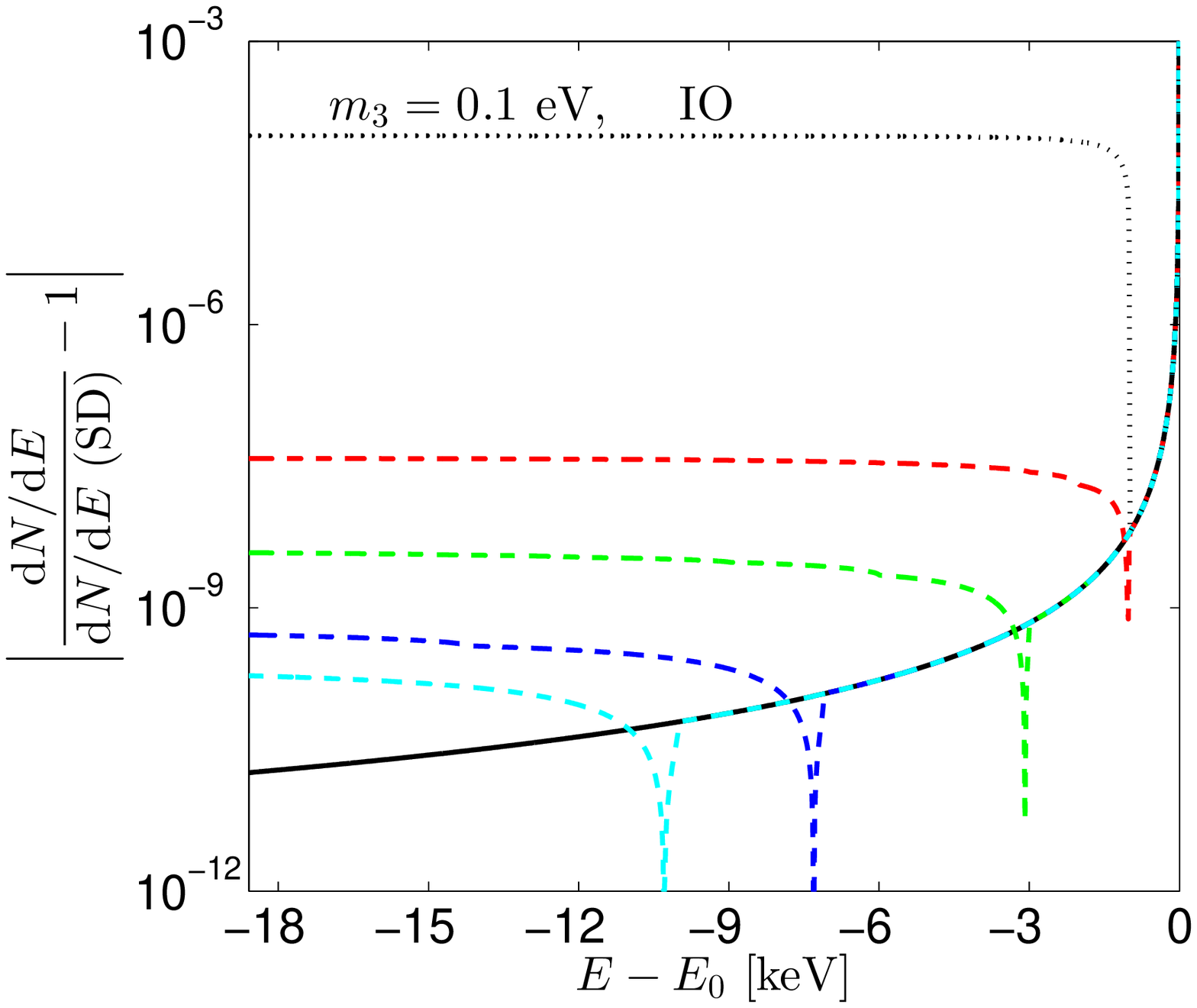}
\caption{\label{fig:R} The ratio of a tritium spectrum with and without including heavy KK sterile states. For the left column we assume a normal mass ordering, while for the right column an inverted mass ordering is assumed. The red, green, dark blue and light blue dashed lines correspond to $R^{-1}=1, ~3, ~7.1, ~10~{\rm keV}$, respectively. For comparison, we also show in black dotted lines an extreme case with $m_s=1~{\rm keV}$ and $\sin^2\theta_s=10^{-4}$.
}
\end{center}
\end{figure}

In Fig.~\ref{fig:Kurie} we illustrate the Kurie
function for $R^{-1}=1~{\rm eV}$, which leads to $\theta^{(1)}_{\rm eff}\simeq 0.14$.
There is only a quite small effect of the second member of the KK tower,
which has mixing $\theta^{(2)}_{\rm eff} \simeq 0.07$.
We point out that the possibility of a KK tower of sterile states
has not been considered for a beta spectrum before.
%
For sterile states heavier than a few eV, the spectrum differences cannot be seen in these kind of plots.
However, a keV sterile neutrino manifests itself in the shape of the beta decay spectrum. At an energy $E=E_0 - m_s$ tritium beta decay into sterile neutrinos of mass $ m_s$ is kinematically allowed and a kink shows up in the spectrum around this energy. We therefore consider the ratio of a tritium spectrum with and without extra dimensions, and illustrate the results in Fig.~\ref{fig:R}.
The kinks, in particular the first one, in the spectrum can be clearly seen from the plot. The effect from the second KK mode appears around $E-E_0 \simeq 2R^{-1}$, which is however not nearly as significant
as that for the first KK threshold. In our framework a nearly degenerate active neutrino spectrum could lead to larger mixing, and more promising signatures in beta decay experiments. For comparison, we also plot an extreme case for $m_s=1~{\rm keV}$ and $\sin^2\theta_s=10^{-4}$, in which the standard and sterile neutrino polluted spectra are well separated.

\begin{figure}[h]
\begin{center}\vspace{0.cm}
\includegraphics[width=.45\textwidth]{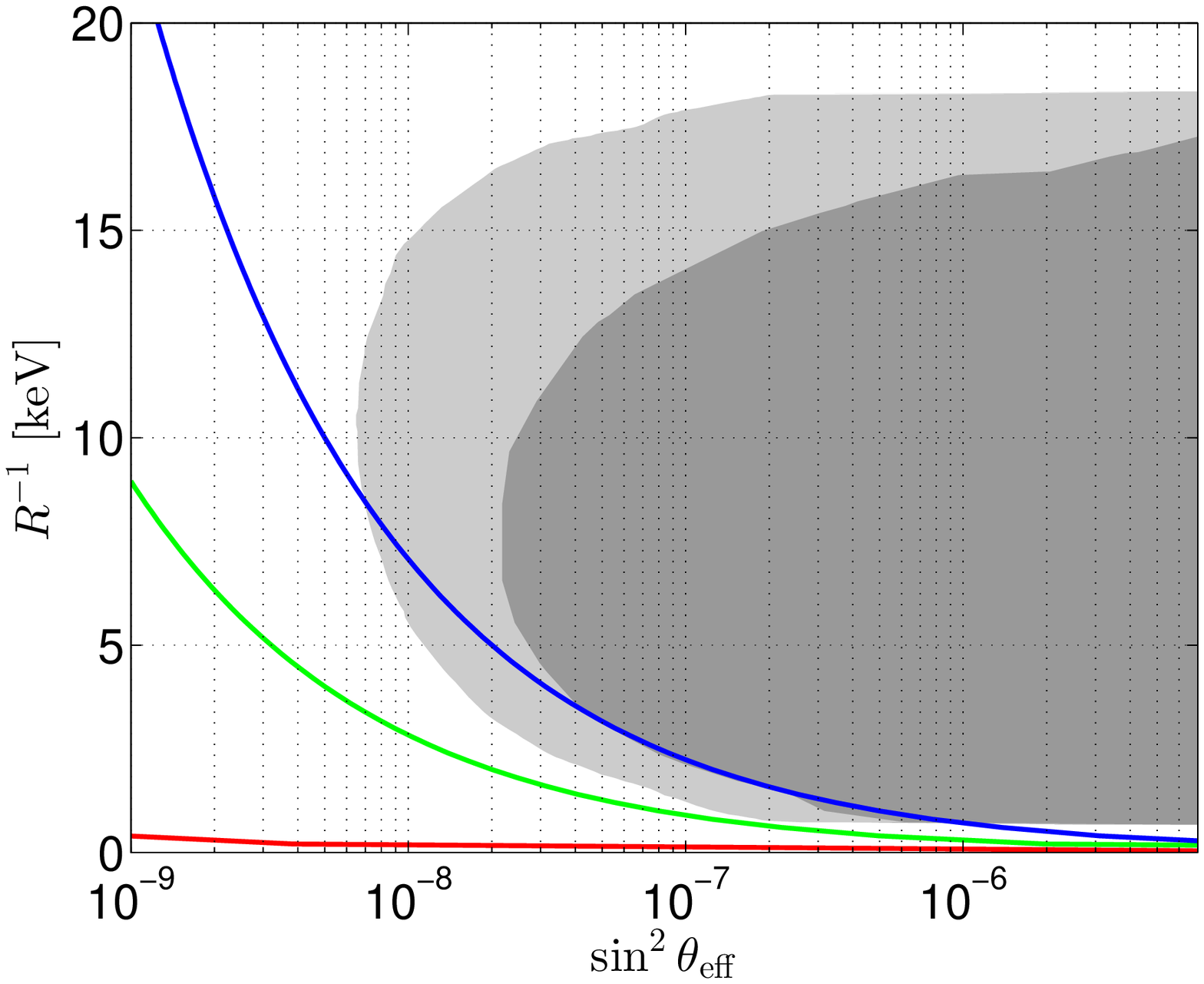}\hspace{5mm}
\includegraphics[width=.45\textwidth]{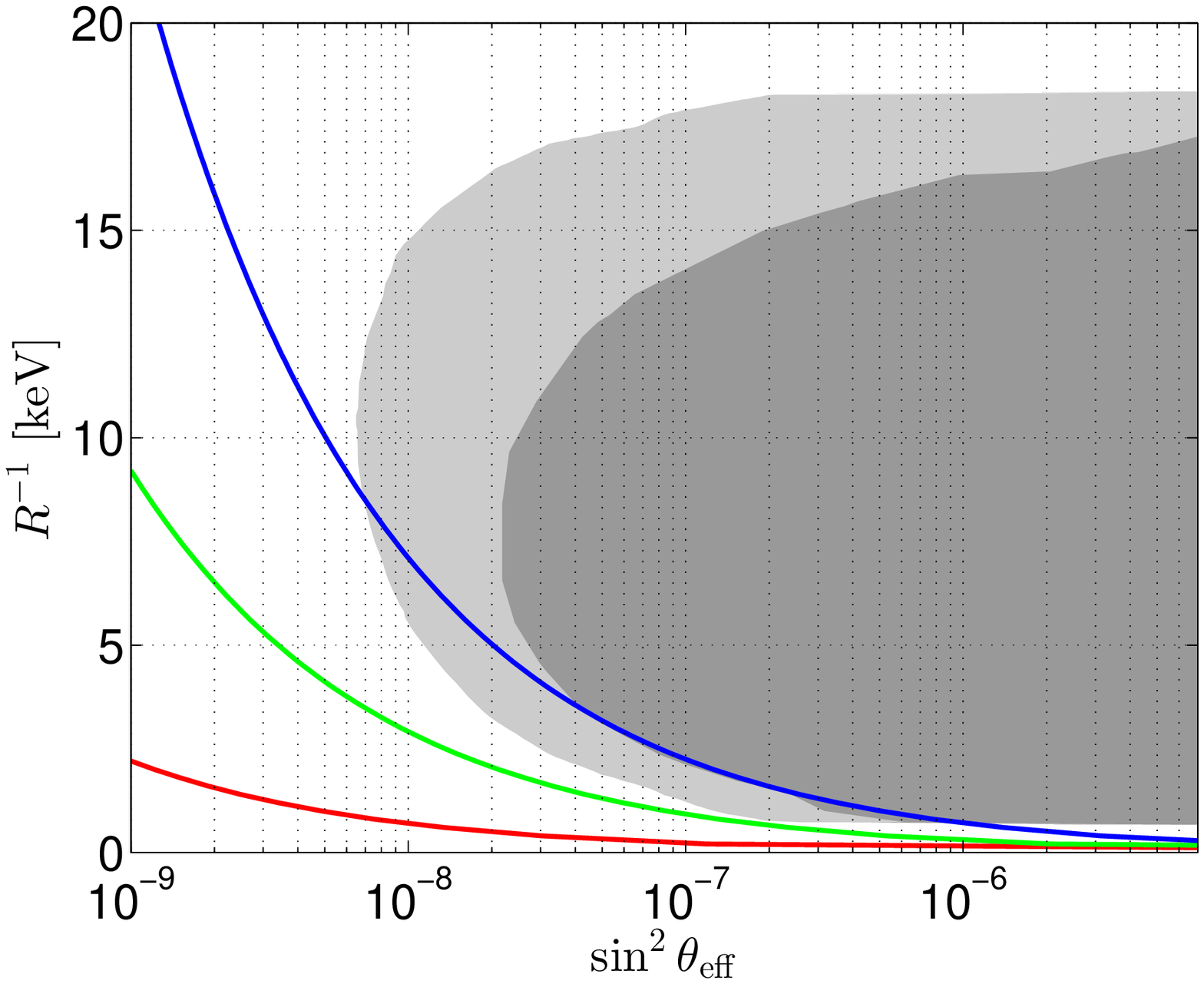}
\caption{\label{fig:KATRIN} Predictions compared to the sensitivity of a modified
KATRIN. Red, green and blue curves correspond to $m_1=0$,
$0.2~{\rm eV}$ or $0.5~{\rm eV}$, ($m_3$ for the inverted hierarchy
case [right plot]). The shaded areas indicate the exclusion regions of a modified KATRIN
at $90\%$~C.L.\ \cite{SM}. }
\end{center}
\end{figure}

While the presence of a KK tower in the beta spectrum has not yet been investigated,
studies indicate that a modified KATRIN setup with access to the full energy spectrum
can reach a sensitivity of a mixing angle down to $\sin^2\theta \sim 10^{-8}$ for a 
single keV sterile neutrino \cite{SM}. 
In Fig.~\ref{fig:KATRIN}, using only the first KK mode,
we illustrate the parameter region of our framework together with the sensitivity
of a modified KATRIN experiment to a keV sterile neutrino \cite{SM}.

\section{Conclusion}
\label{sec:summary}

In this work we have studied an extra dimension model for Dirac neutrinos
with the lightest KK modes being sterile neutrinos around the eV or keV scale.
We pointed out that in the framework under study the active-sterile mixing parameters are directly
connected to active neutrino mixing parameters. In case of eV-scale sterile neutrinos,
the predicted active-sterile mixing could lead to sizable $\nu_e \to \nu_s$ transitions,
accounting for the observed short-baseline neutrino anomalies. If the radius of the
extra dimension is of order keV, the lowest KK modes could be warm dark matter candidates.
We investigated the beta spectrum in KATRIN-like experiments for eV- and keV-scale KK modes,
pointing out in particular the presence of multiple kinks due to the different KK modes.

\section*{Acknowledgment}
H.Z.\ thanks NORDITA for hospitality during the completion of this work. This work was supported by the Max Planck Society in the project MANITOP.
\bibliography{bib}

\end{document}